%% file: EdgeSpinChains.tex
\documentclass[prl,aps,longbibliography,floats,superscriptaddress,twocolumn]{revtex4-1}

\usepackage{ae}
\usepackage[T1]{fontenc}
\usepackage[utf8]{inputenc}
\usepackage{amsmath,amssymb,amsthm,amsbsy,bm}
\usepackage{graphicx,epsfig,epstopdf,subfigure}
\usepackage{dsfont,verbatim,varwidth,dcolumn}
\usepackage[usenames,dvipsnames]{xcolor}
\usepackage{color,soul,subfiles}
\usepackage[colorlinks]{hyperref}

\def\ua{{\uparrow}}
\def\da{{\downarrow}}

\newcommand{\vsr}{\vec{r}}
\newcommand{\vR}{\vec{R}}

\newcommand{\cdo}[1]{c^{\dagger}_{#1}}
\newcommand{\co}[1]{c^{ }_{#1}}

\makeatother

\begin{document}

\subfile{MainDraft.tex}

\onecolumngrid
\clearpage

\subfile{Supplemental.tex}

\end{document}

%% file: MainDraft.tex
\title{Edge Reconstruction of a Time-Reversal Invariant Insulator: Compressible-Incompressible Stripes}

\author{Udit Khanna}
\altaffiliation{Present Address: Department of Physics, Bar-Ilan University, Ramat Gan 52900, Israel}
\email{uditkhanna10@gmail.com}
\affiliation{Department of Condensed Matter Physics, Weizmann Institute of Science, Rehovot 76100, Israel}

\author{Yuval Gefen}
\email{yuval.gefen@weizmann.ac.il}
\affiliation{Department of Condensed Matter Physics, Weizmann Institute of Science, Rehovot 76100, Israel}

\author{Ora Entin-Wohlman}
\email{orawohlman@gmail.com}
\affiliation{School of Physics and Astronomy, Tel Aviv University, Tel Aviv 6997801, Israel}

\author{Amnon Aharony}
\email{aaharonyaa@gmail.com}
\affiliation{School of Physics and Astronomy, Tel Aviv University, Tel Aviv 6997801, Israel}

\begin{abstract}

Two-dimensional (2D) topological electronic insulators are known to give rise to gapless edge modes, which 
underlie low energy dynamics, including electrical and thermal transport. This has been thoroughly 
investigated in the context of quantum Hall phases, and time-reversal invariant 
topological insulators. Here we study the edge of a 2D, topologically trivial insulating phase, as a
function of the strength of the electronic interactions and the steepness of the confining potential. 
For sufficiently smooth confining potentials, alternating compressible and incompressible stripes appear at the edge. 
Our findings signal the emergence of gapless edge modes which may give rise to finite 
conductance at the edge. This would suggest a novel scenario of a nontopological metal-insulator transition 
in clean 2D systems. The incompressible stripes appear at commensurate fillings and may exhibit broken translational 
invariance along the edge in the form of charge density wave ordering. 
These are separated by structureless compressible stripes. 
\end{abstract}

\maketitle

{\it Introduction}.--The edge of two-dimensional (2D) topological insulators is underlined by gapless chiral (or helical) 
modes~\cite{Halperin1982,Wen1990,Wen1992,WenBook}.
Their number and directionality are constrained by the bulk-edge correspondence~\cite{WenBook}. 
Varying the steepness of the confining potential at the edge and the strength of the electron-electron interaction 
may give rise to a sequence of quantum phase transitions, also known as ``edge reconstruction.''  
These are marked by the emergence of additional chiral edge modes, abiding by
the bulk-edge correspondence. Edge reconstruction has been discussed in the context of the 
integer~\cite{CSG1992,Dempsey1993,ChamonWen,Sondhi_PRL_96,FrancoBrey97,KunYangIQHS,Switching2017,Ganpathy21,Nu12020} and 
fractional~\cite{MacDonald_PRL_90,Johnson_PRL_91,MacDonald_JP_93,Meir93,
KFP1994,KaneFisher_PRB_95,Ganpathy_PRB_03,KunYang_PRL_02,KunYang_PRB_03,KunYang03,WMG-13,Jain2014,Yang2021} quantum Hall 
(QH) phases, and has been proposed as a viable mechanism for time-reversal invariant topological 
insulators~\cite{Yuval2017,Rosenow2021}. For all these instances of edge reconstruction, driven by charging 
and/or exchange effects, the bulk state remains untouched. 

\begin{figure}[t]
  \centering
  \includegraphics[width=0.49\columnwidth]{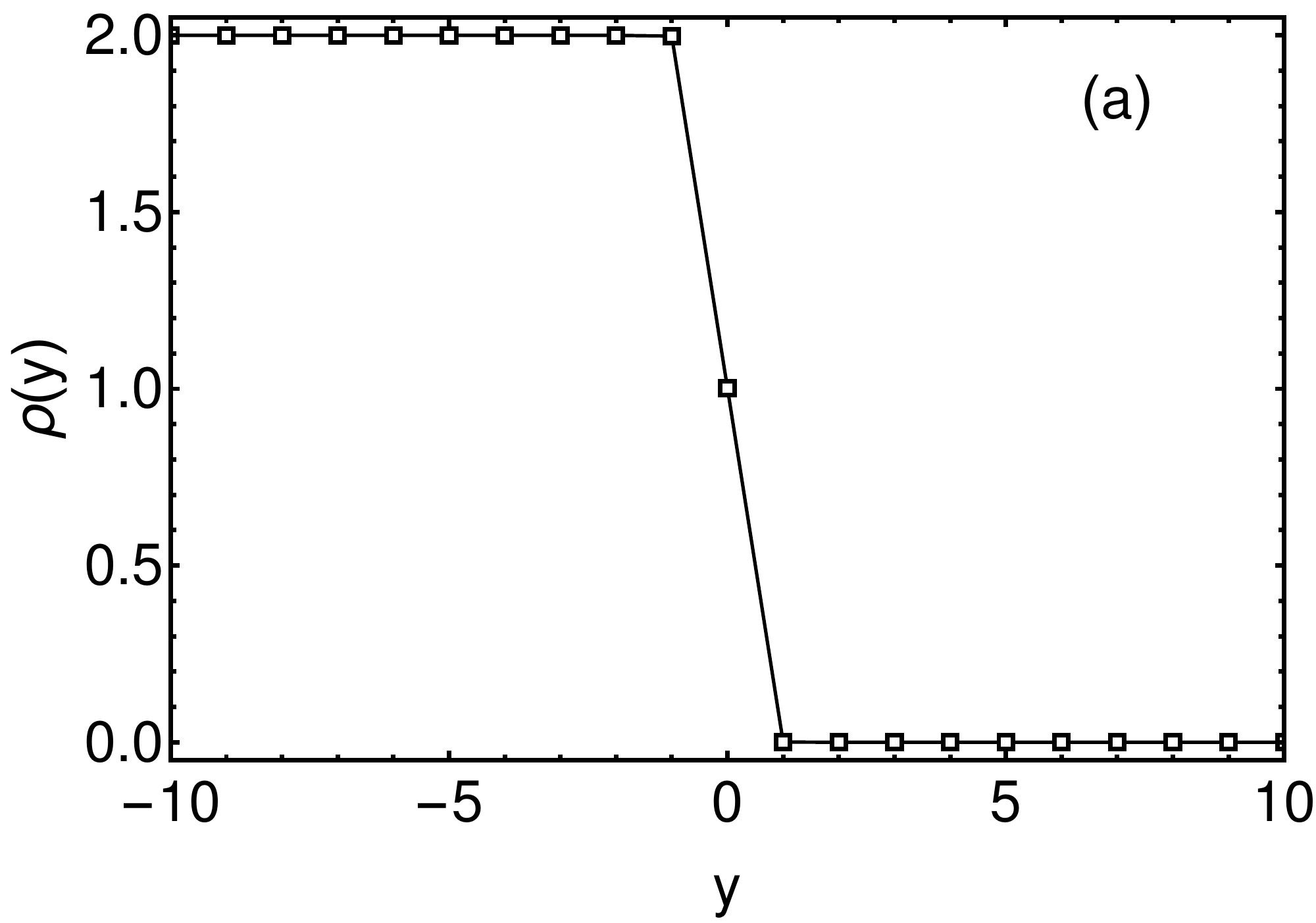}
  \includegraphics[width=0.49\columnwidth]{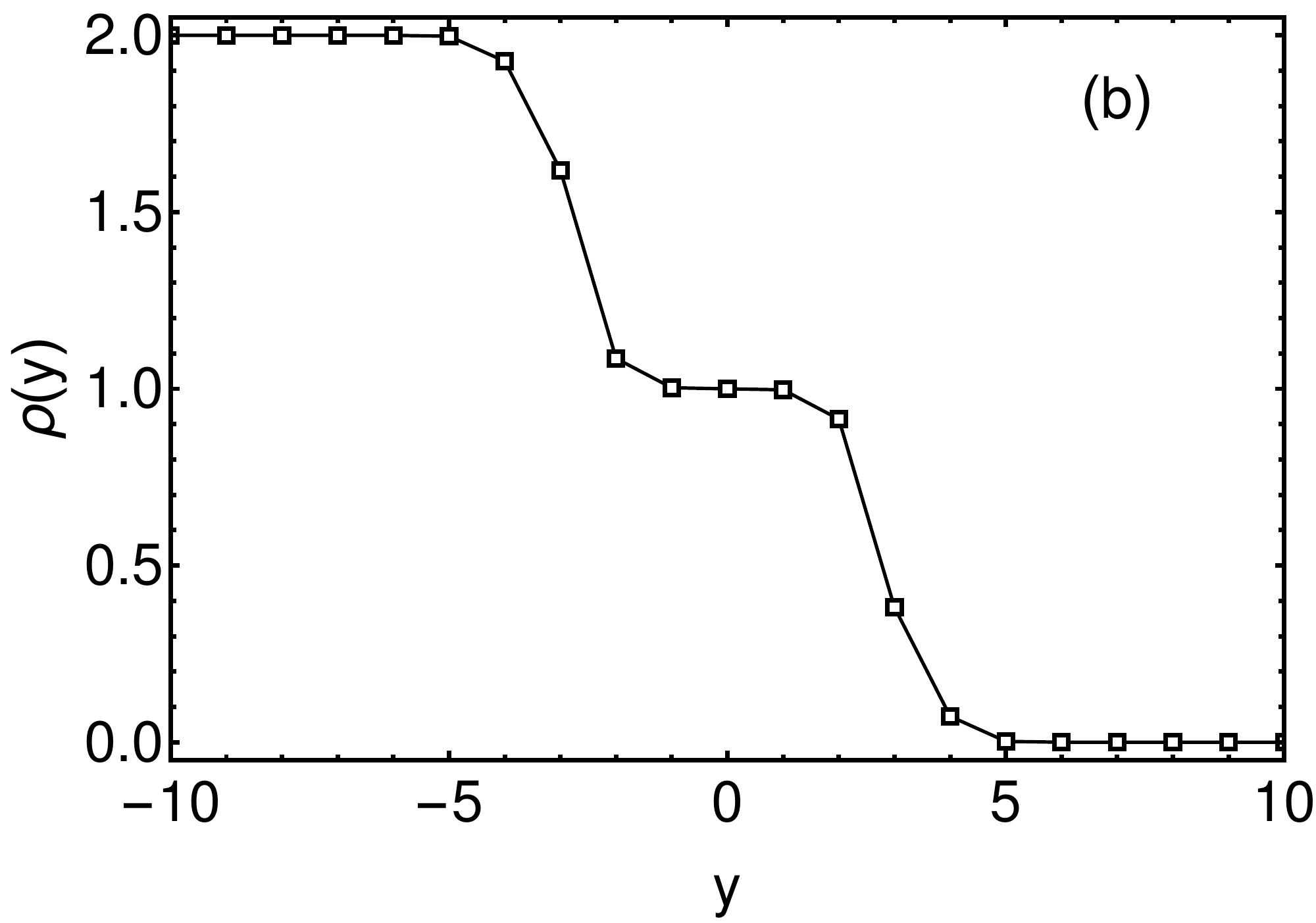}
  \includegraphics[width=0.49\columnwidth]{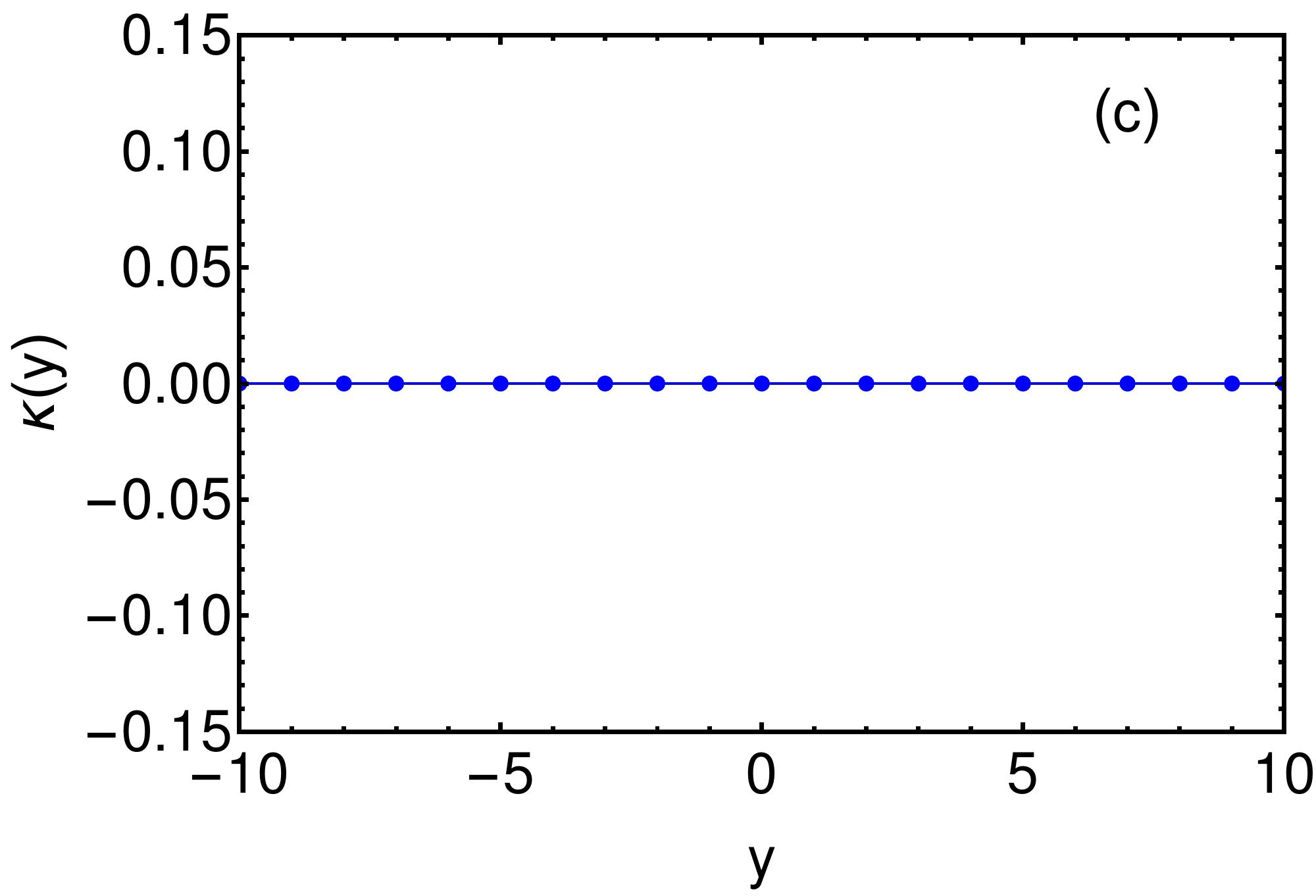}
  \includegraphics[width=0.49\columnwidth]{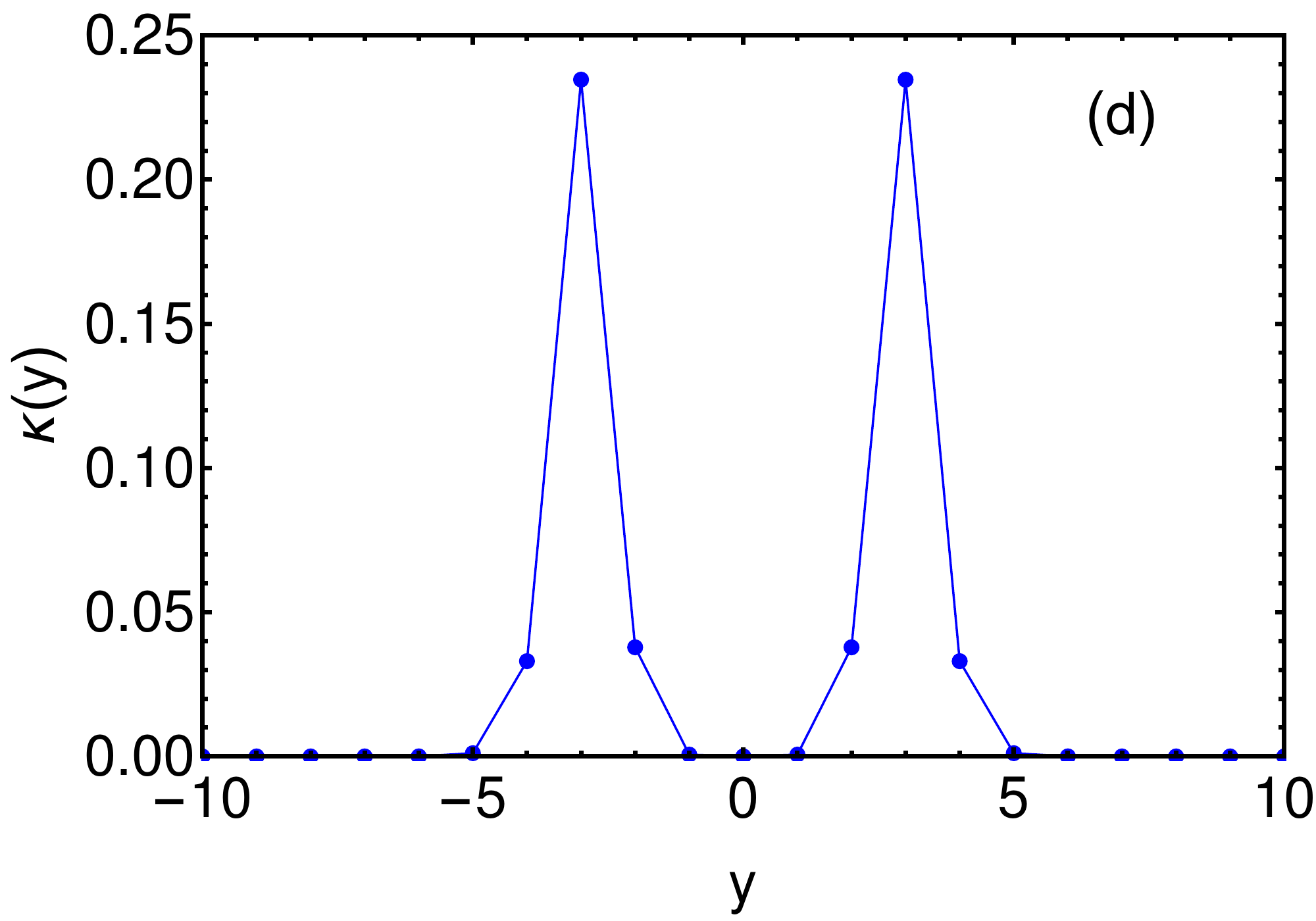}
  \caption{ 
  Changes in the edge structure driven by variations of slope of the confining potential ($w^{-1}$). 
  (a)--(b) Average charge density as a function of the distance
  from the edge ($y$) for (a) sharp ($w = 0.5$) and (b) smooth ($w = 8.5$) confining potentials.  
  Regions at large negative (positive) $y$ are part of the insulator (vacuum) and are doubly occupied
  (empty). (c)--(d) Local compressibility (defined in the text) 
  of the states presented in panels (a)--(b) respectively. As seen smoothening
  the potential leads to the formation of compressible stripes 
  and an incompressible half-filled plateau. 
  The parameters used here are $t = V = 1$ and $U = 15$. }
\end{figure}

\begin{figure*}[t]
  \centering
  \includegraphics[width=0.24\textwidth]{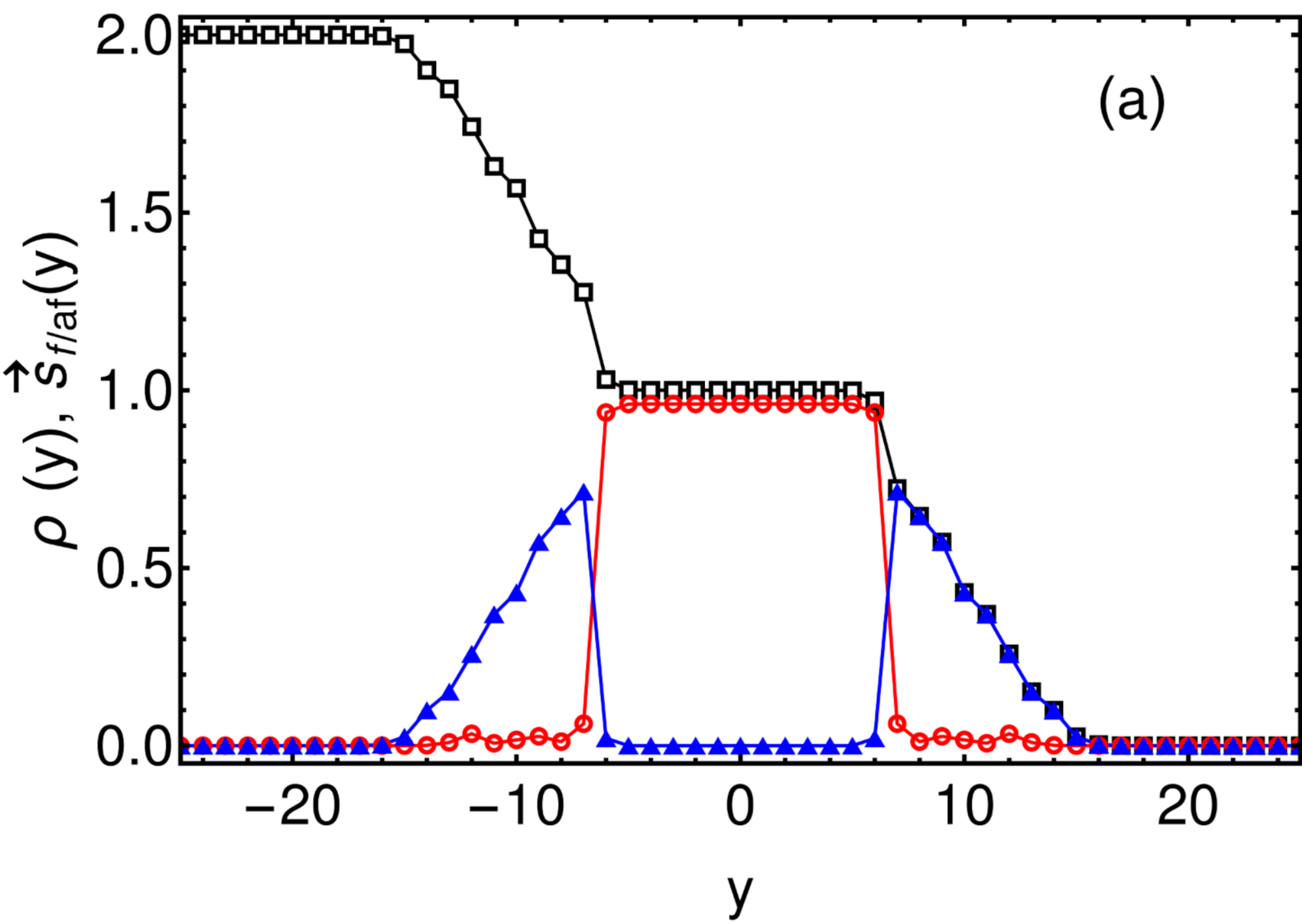}
  \includegraphics[width=0.24\textwidth]{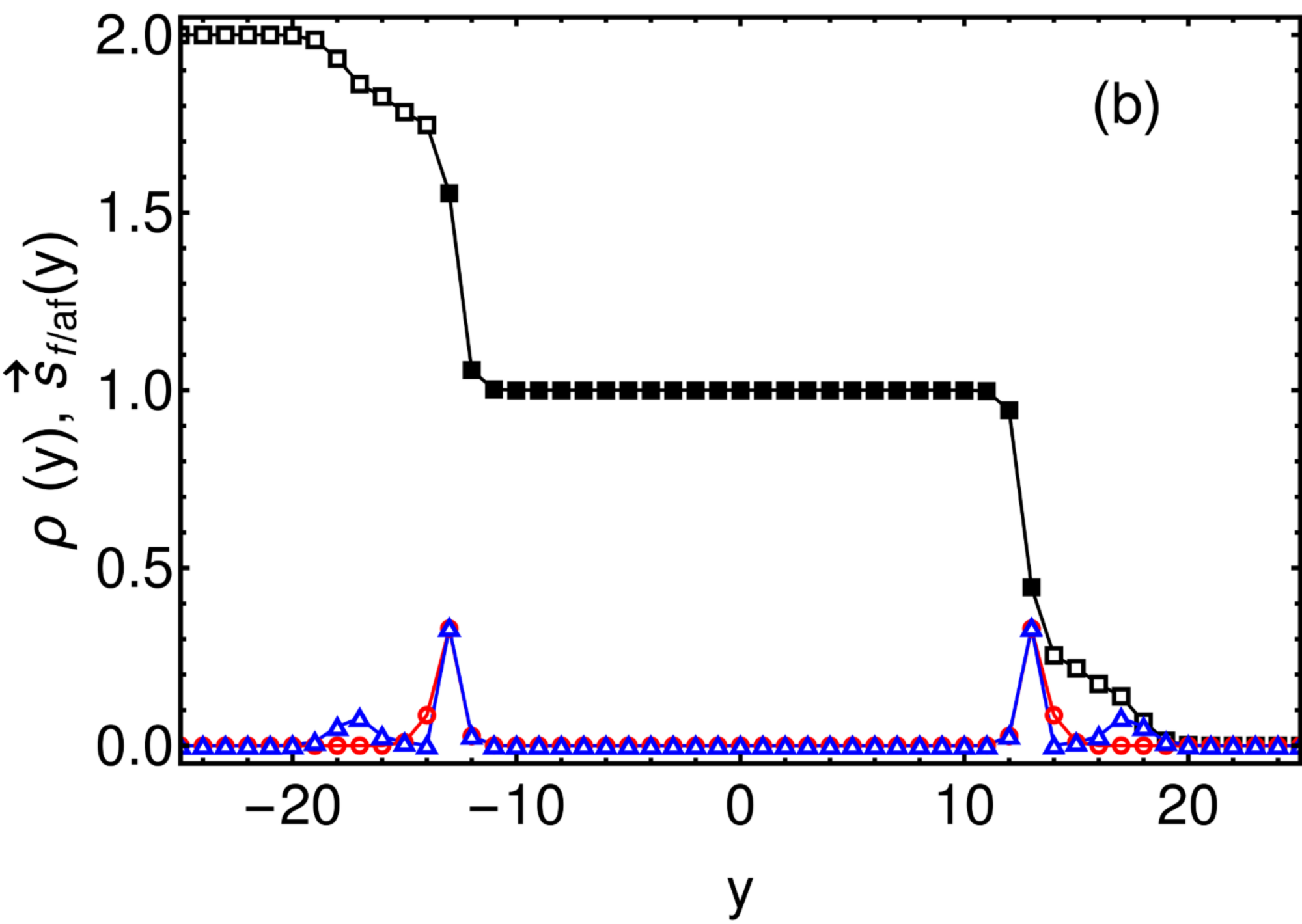}
  \includegraphics[width=0.24\textwidth]{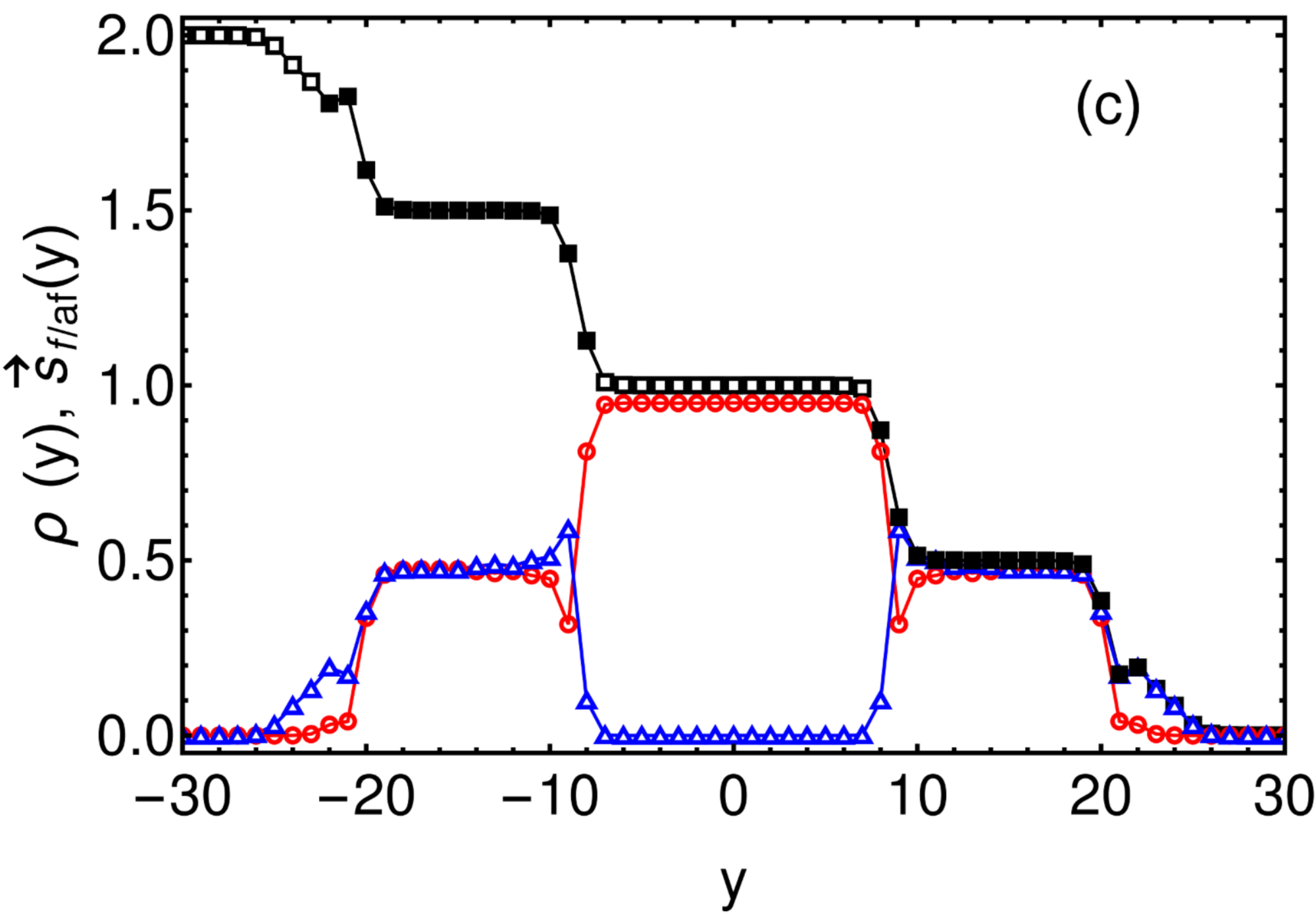}
  \includegraphics[width=0.24\textwidth]{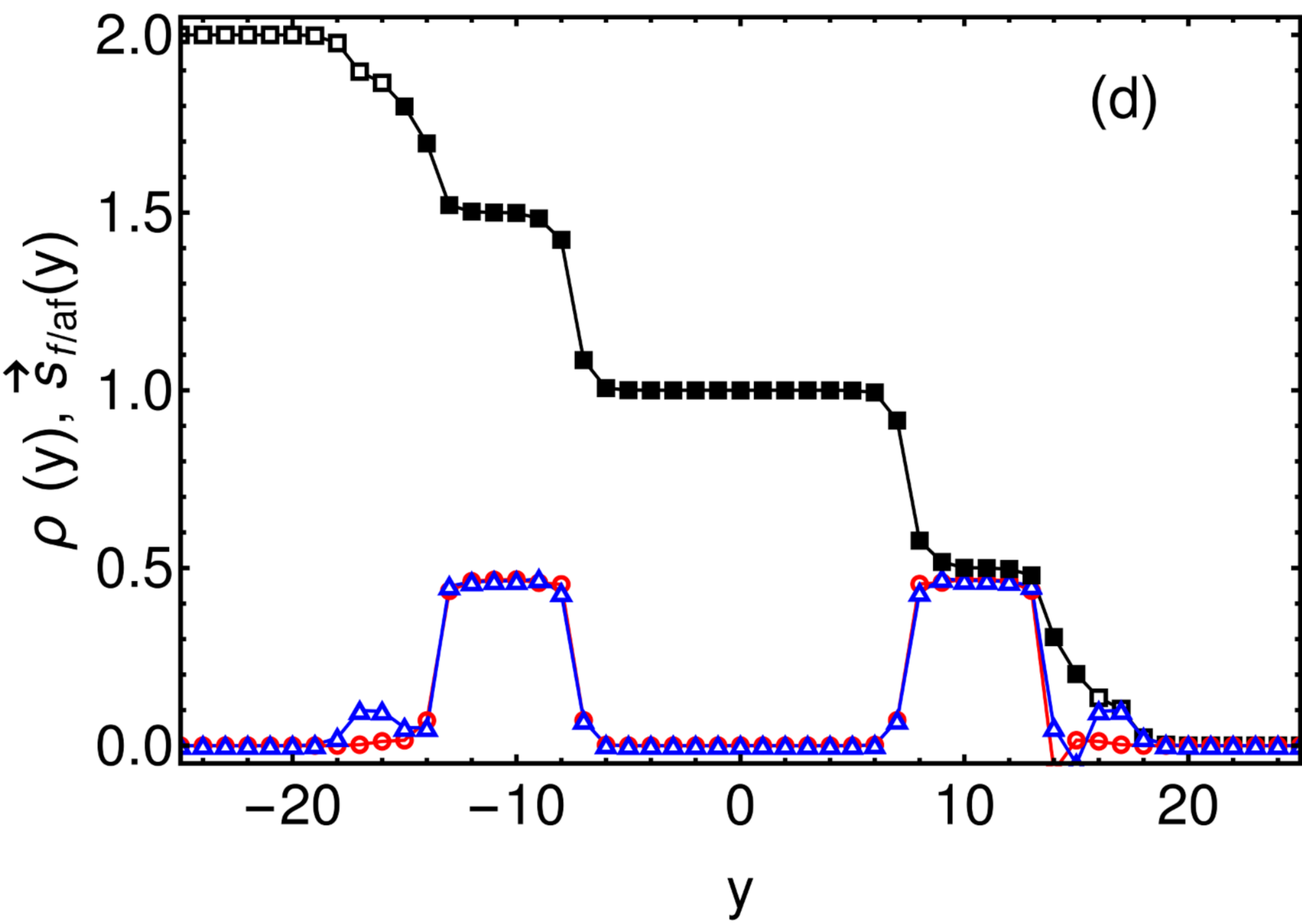}
  \includegraphics[width=0.24\textwidth]{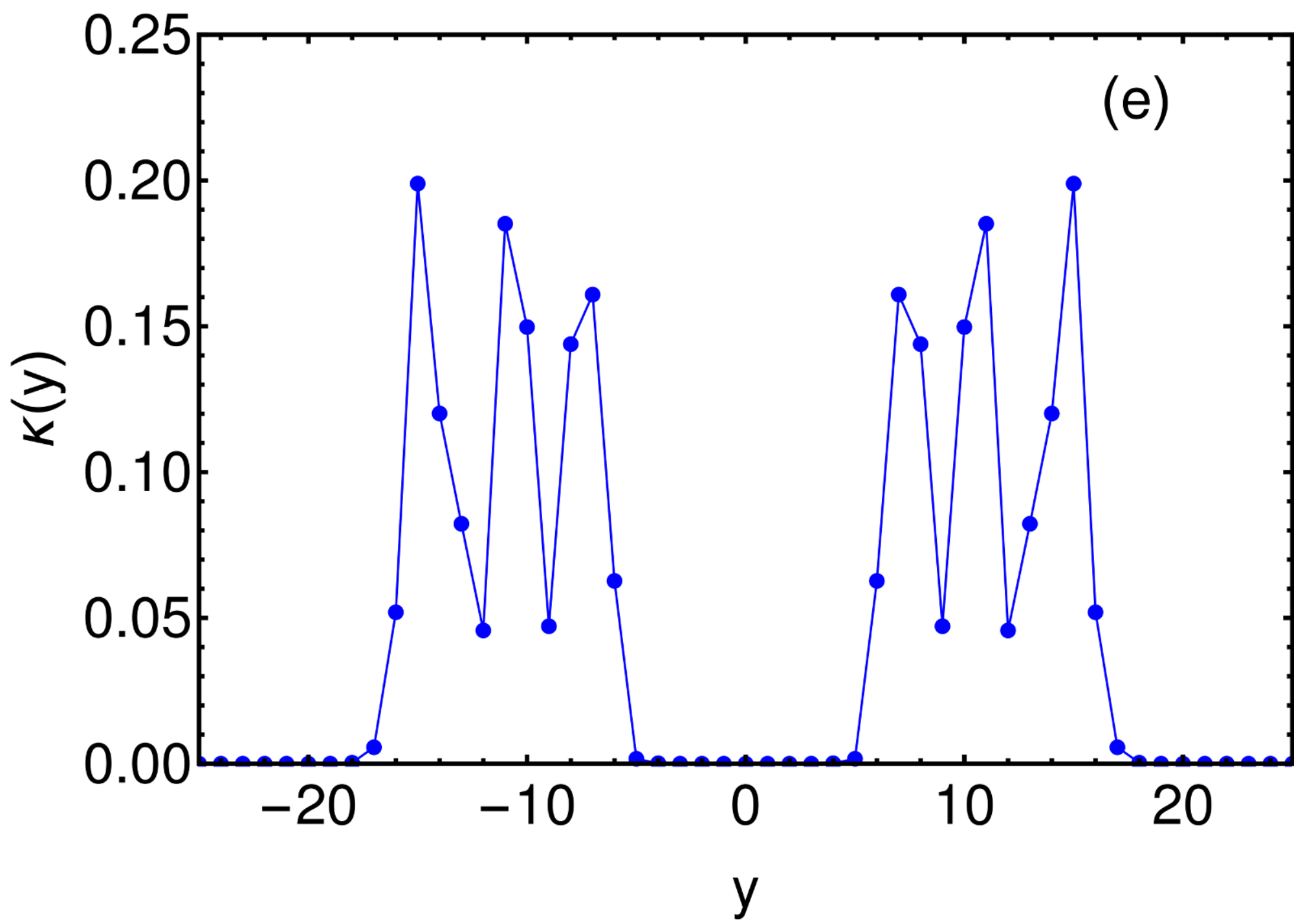}
  \includegraphics[width=0.24\textwidth]{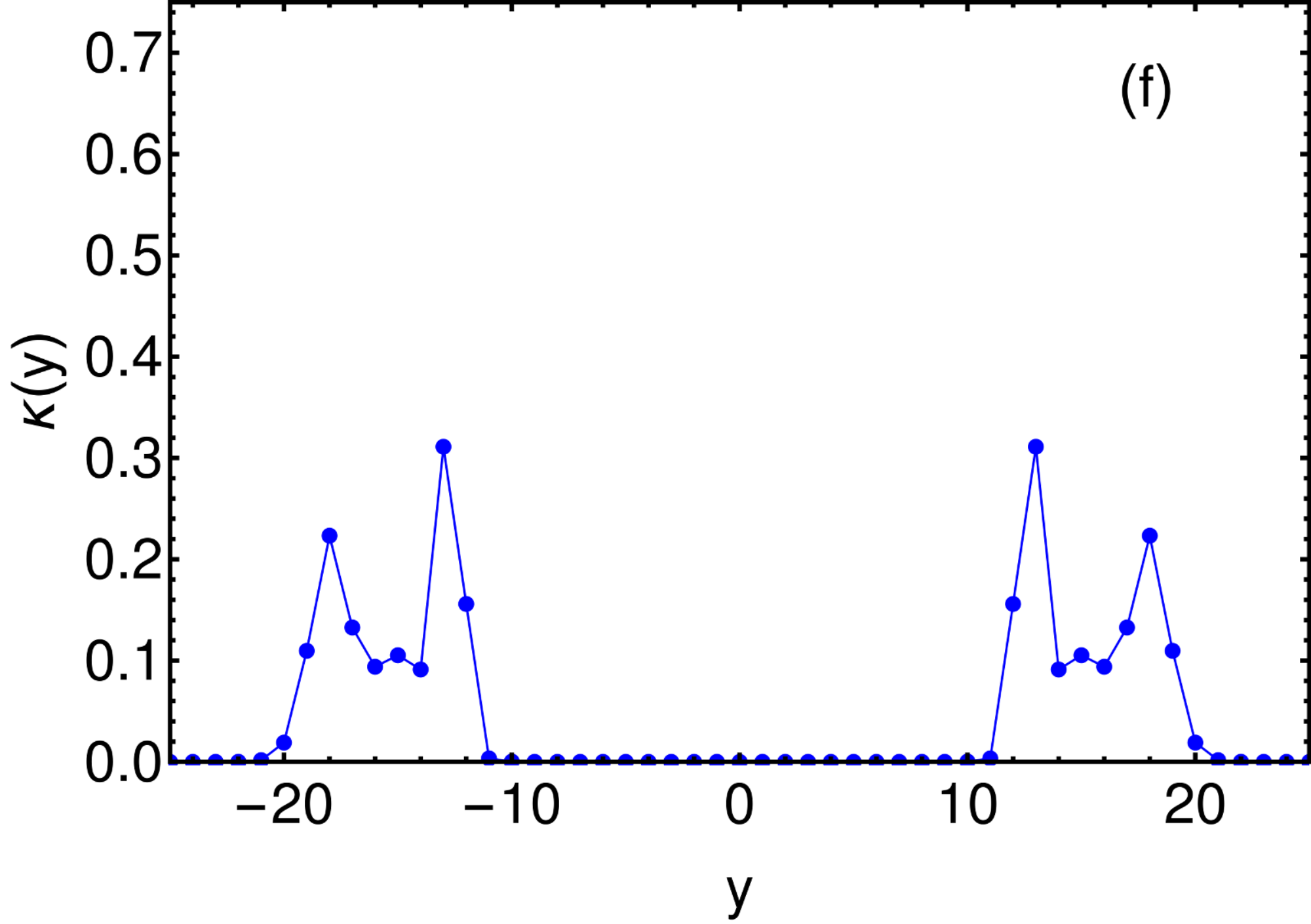}
  \includegraphics[width=0.24\textwidth]{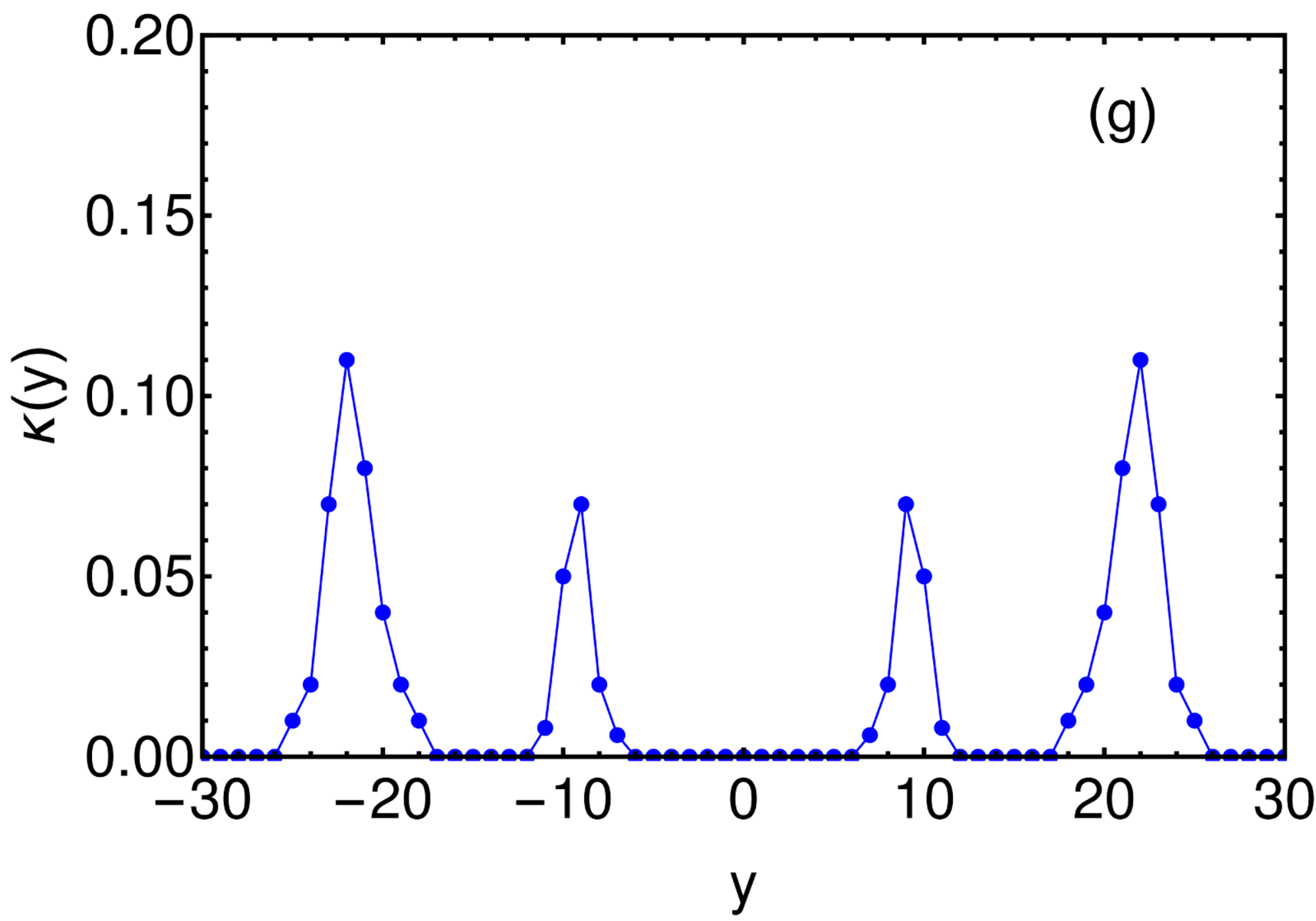}
  \includegraphics[width=0.24\textwidth]{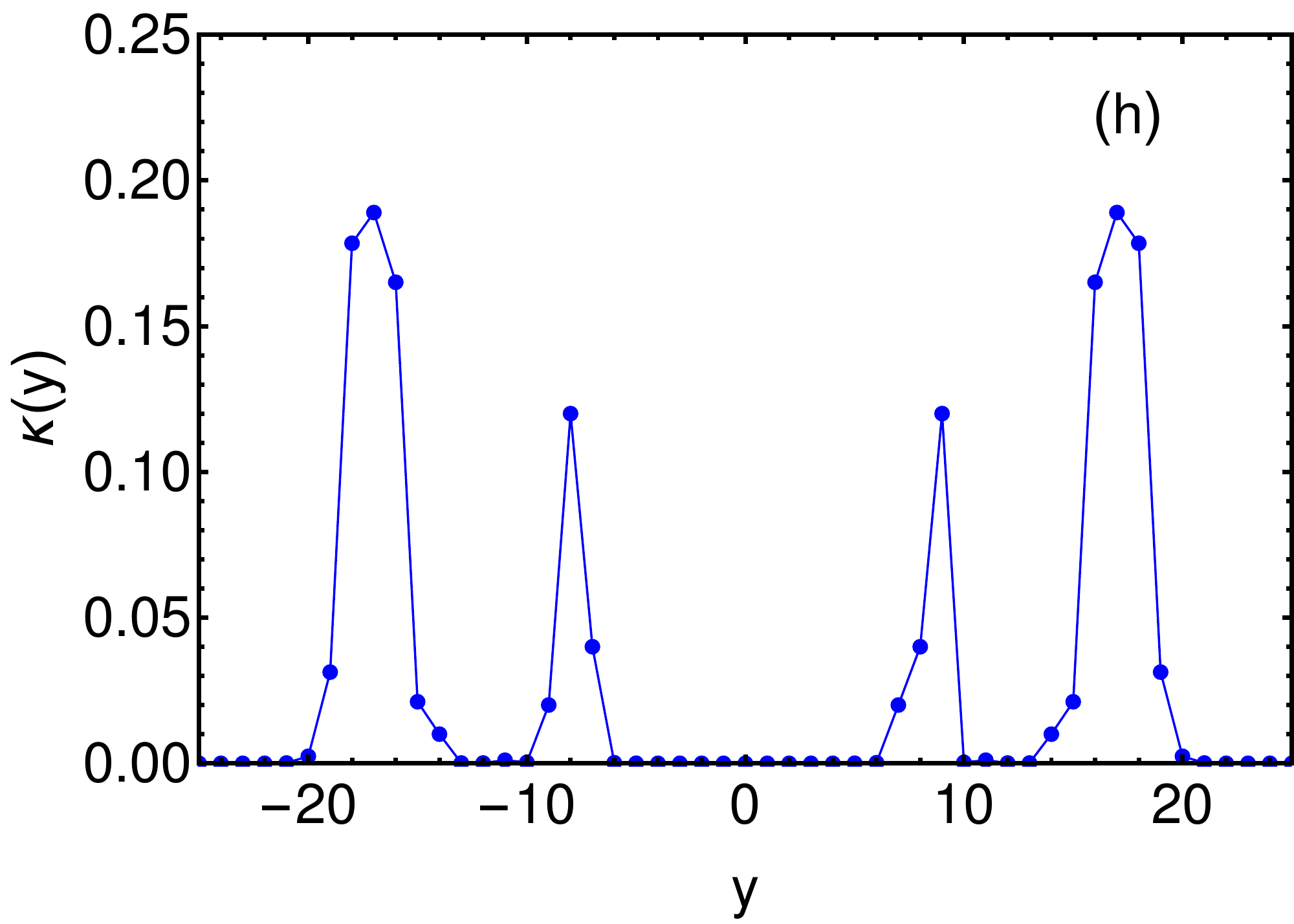}
  \caption{ 
  (a)--(d) The $y$ dependent order parameters (defined in the text) in the four edge structures found for 
  smooth confining potentials. An identical confining potential ($w = 50$) is used for all panels, while the interaction strengths 
  ($U$ and $V$) vary. Moving from (a) to (b), $V$ is increased while $U$ is fixed. Moving from (b) to (d), $U$ is 
  decreased at a constant $V$. Black curves (empty and filled squares) show the average occupation [$\rho(y)$]. 
  Filled [empty] squares denote regions with a CDW of period 2 [a uniform charge density] along the edge 
  (along $x$). The blue [red] line (triangles [circles]) shows the average ferromagnetic (F) [antiferromagnetic (AF)] order parameter. 
  The empty (filled) triangles and circles denote magnetic order along the $s_z$ ($s_x$) direction. 
  This direction is chosen arbitrarily for the moments on the half-filled stripe.  
  All edge structures feature a central plateau at half filling ($\rho = 1$). This central plateau 
  has (a) AF order (in the limit $U \gg V \gg t$) or (b) CDW order (in the limit $V \gg U$). (c),(d) Additional plateaus 
  at $\rho = 1/2$ and $3/2$ can appear when both $U$ and $V$ are large. These plateaus feature a CDW order 
  of period 2. Since the (average) spin on each partially occupied site in these regions is equal, the average F
  and AF order parameters are identical. The additional plateaus do not change the order on the central plateau which can 
  be either (c) AF or (d) CDW. (e)--(h) The local compressibility in the states depicted in panels 
  (a)--(d) respectively. The presence of alternating compressible and incompressible stripes is clear. 
  The values of $U$ and $V$ used here are shown in Fig.~4. }
\end{figure*}

Here, we consider the boundary of a clean, topologically trivial, 2D insulator at zero magnetic field. 
The edge structure of such a system may depend crucially, not only on the microscopic bulk Hamiltonian and 
the electronic interactions, but also on the steepness of the edge confining potential.
Considering a lattice model to study the boundary, we indeed find,
for a sufficiently smooth confining potential, the emergence of alternating compressible and incompressible 
stripes at the edge (Fig.~1). The widths of the stripes decrease with the increase of slope of the potential. 
Each incompressible stripe has a constant filling [plateaus at half filling in Fig.~1(b), Figs.~2(a) and 2(b) or quarter filling in Figs.~2(c) and 2(d)].
By contrast, in the compressible stripes, the average filling varies smoothly and monotonically
as the edge is approached; this appears in the regions surrounding the incompressible plateaus [Figs.~1(b) and 1(d)]. 
Furthermore, the low energy tunneling density of states vanishes (remains finite) in the
incompressible (compressible) stripes (Fig.~5).
Since metallic behavior is accompanied by finite compressibility~\cite{CSG_Conductivity,Codben99,Ilani_Compressibility,
Kaan_2003,Armagnat_2020}, we conjecture that the compressible stripes have finite conductance. 
This would represent a metal-insulator transition at the edge of a clean system, driven by the slope of the boundary confining potential.

The edge of 2D topological phases of matter, which supports a discrete (and small) number of chiral modes, may naturally be 
described as a quasi one-dimensional (1D) system. 
In such systems the competition between the kinetic energy and the Coulomb repulsion as well as the interplay of electrostatic and 
exchange interactions give rise to a multitude of electronic phases of matter. 
For strictly 1D systems, modeled as Luttinger liquids, a number of phases 
(marked by power-law correlations of charge, spin, and superconducting fluctuations) may emerge~\cite{Solyom,Emery,Giamarchi}.
For lattice-based models, such as the extended Hubbard model, truly long-range charge-ordered phases may 
emerge with commensurate electronic filling factors~\cite{Giamarchi,Schulz90,Voit92,Mila93,Mila94}.
Generalizations of 1D systems to ladders with two~\cite{Giamarchi,Ladder1,Ladder2,Ladder3,CDW-Ladder,CDW-Ladder2} and 
three~\cite{3LL1,PRBChinese2016} coupled chains have been reported. Lattice-based models of ladders 
yield similar commensurate charge-ordered phases~\cite{Giamarchi,CDW-Ladder,CDW-Ladder2}. 

By contrast, here the edge comprising compressible and incompressible stripes has a finite width, consisting of 
a large number of parallel 1D {\it chains}. 
This renders an analysis based on bosonization of a multilegged ladder impractical. Instead, we employ
a self-consistent Hartree-Fock (HF) analysis aimed at finding the ground state of the chains, 
coupled through tunneling and electron-electron interactions, as a function of 
the slope of the confining potential and the interaction strength. 
In addition to the nucleation of compressible stripes, our HF analysis results in charge and 
magnetic order in the incompressible stripes for different interaction strengths [Fig.~2].
Noting that a phase with charge density wave (CDW) order breaks a discrete symmetry, a truly long-range
charge ordering is expected to exist in quasi-1D systems~\cite{Giamarchi,CDW-Ladder,CDW-Ladder2}.
Therefore, even though we perform a self-consistent mean-field analysis, our results vis-a-vis charge ordering
may survive the inclusion of quantum fluctuations. 
The presence of alternating compressible and (ordered) incompressible 
stripes at the edge of a trivial insulator is the main result of this Letter.

We note that Ref.~\cite{CSG1992} reported the presence of alternating compressible-incompressible stripes at the edge 
of QH phases. In that case the incompressible stripes appear in the regions with a quantized filling factor. We stress 
that our results differ from Ref.~\cite{CSG1992} since we consider a topologically trivial phase in the absence of a 
magnetic field. 

\begin{figure}[t]
  \centering
  \includegraphics[width=0.72\columnwidth]{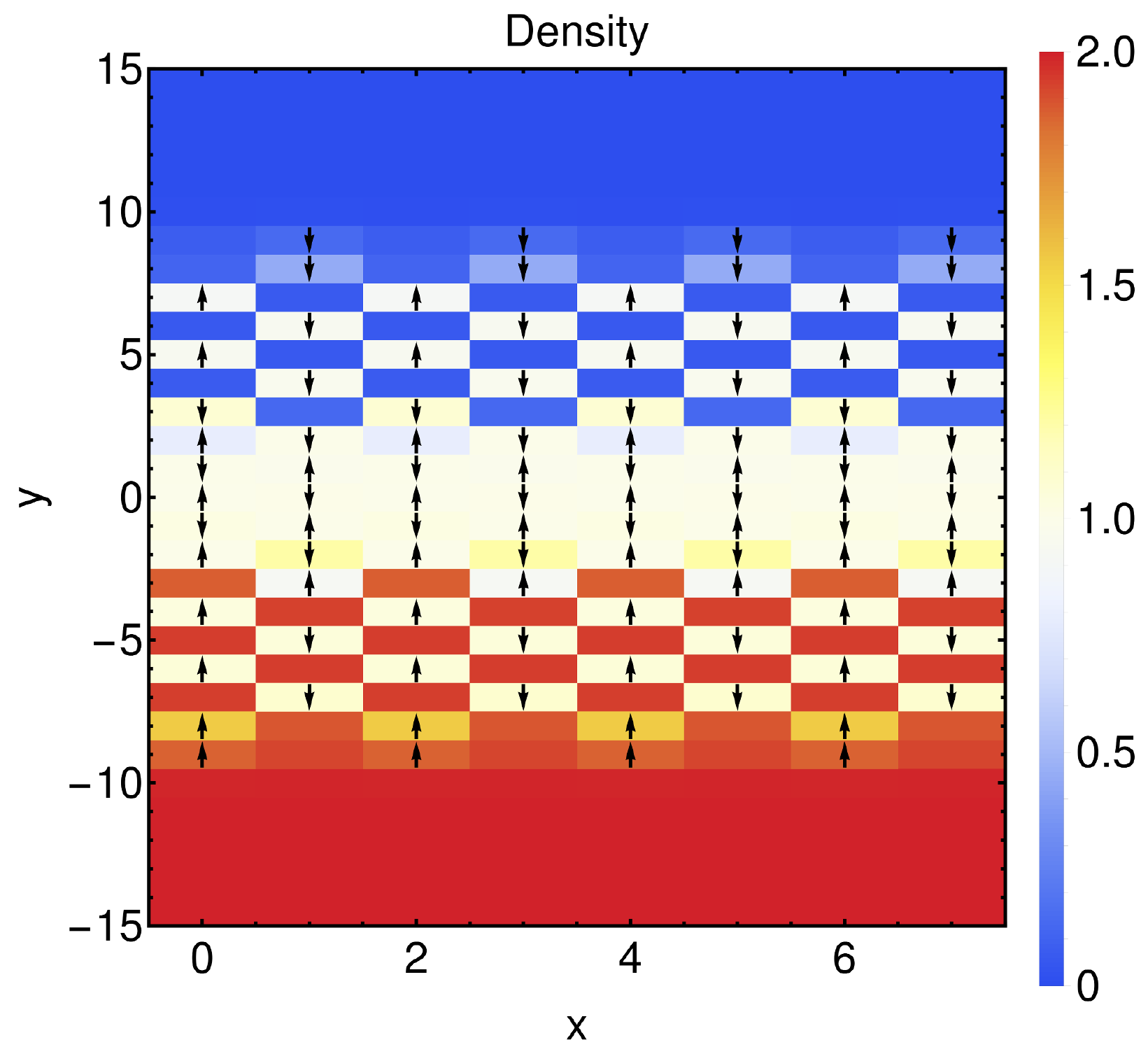}
  \caption{ Ordered stripes at the edge of a 2D insulator. The color map shows the average occupation 
  at each site. The black arrows represent the average spin (along $s_z$) at each site. 
  Only the first eight sites along the edge ($x$ direction) are shown. 
  The region around $y = 0$ features single occupied sites with 
  AF ordering [the central plateau in Fig.~2(c)].  
  The surrounding regions host a CDW (marked by the alternating colors),    
  forming the quarter-filled plateaus in Fig.~2(c). 
  This is the result for 30 chains (along $y$ direction) with $w = 12$. 
  The parameters $U$, $V$, and $t$ are the same as those used for Fig.~2(c). }
\end{figure}

{\it Basic setup}.--We analyze the edge of a 2D insulator using the single band extended Hubbard model on a square lattice (lattice spacing 1)
with a confining potential perpendicular to the edge (along $y$). 
We assume a large bulk band gap so that the conduction band plays no role.
The Hamiltonian of the system is $H = H_{o} + H_{ee}$ where $H_{ee}$ is the Coulomb repulsion and $H_{o} = \sum_{\vsr, \sigma} [-\mu + V_{c}(y)] 
c^{\dagger}_{\vsr, \sigma} c^{}_{\vsr, \sigma} -t \sum_{\langle \vsr, \vR \rangle, \sigma} 
[ c^{\dagger}_{\vsr, \sigma} c^{}_{\vR, \sigma} + \, \text{H.c.} ] $. 
Here $\sigma$ are the spin indices (quantized along an arbitrary direction) and $t$ is the isotropic 
hopping integral between nearest neighbors. The chemical potential $\mu$ is chosen such that 
the bulk sites (at large and negative $y$) are doubly occupied. We model the confining potential as a linear function,
\begin{align}
  V_{c} (y)= \Bigg\{\begin{array}{cc}
    0 & y < -w   \\
    V_{0} + \frac{V_{0}}{w} y & y \geq -w \end{array}. 
\end{align}
Here $V_{0}$ is chosen such that the average occupation 
[$\rho(y) = \sum_{x,\sigma} \langle c^{\dagger}_{\vsr, \sigma} c^{}_{\vsr, \sigma} \rangle / L_{x}$] 
of the chain at $y = 0$ is $1$. The parameter $w^{-1}$ is the slope of the 
edge potential. This potential is very steep at $w \sim 0$ and becomes smoother as $w$ increases.
Since any sufficiently smooth function can be linearized
around the chemical potential, we do not expect our results to depend on the specific form of $V_c$.
We restrict the range of electronic interactions to nearest neighbors (nn), i.e.,
$H_{ee} = U \sum_{\vsr} n_{\vsr, \ua} n_{\vsr, \da} + V \sum_{\langle \vsr, \vR \rangle}  n_{\vsr} \, n_{\vR}$ 
where $n^{}_{\vsr, \sigma} = c^{\dagger}_{\vsr, \sigma} c^{}_{\vsr, \sigma}$ is the (spin-resolved) number operator and 
$n^{}_{\vsr} = \sum_{\sigma} n^{}_{\vsr, \sigma}$. 
$U$ ($V$) is the on site (isotropic nn) interaction strength. 

We impose periodic boundary conditions (PBC) along the edge ($x$ direction) and open boundary conditions along $y$.   
Performing a Fourier transform along $x$, 
we treat the two-body terms in the HF approximation so that the many-body ground state
is replaced by a variational Slater determinant. The variational parameters are the one-body averages 
$\Delta_{\sigma \sigma'} (k, q, y, \delta) = \langle c^{\dagger}_{k, y, \sigma} c^{}_{k+q, y+\delta, \sigma'} \rangle$.
The HF ground state (which minimizes the variational energy) is found through an iterative procedure carried
out until self-consistency is achieved~\cite{Supplemental}. 

\begin{figure}[t]
  \centering
  \includegraphics[width=0.65\columnwidth]{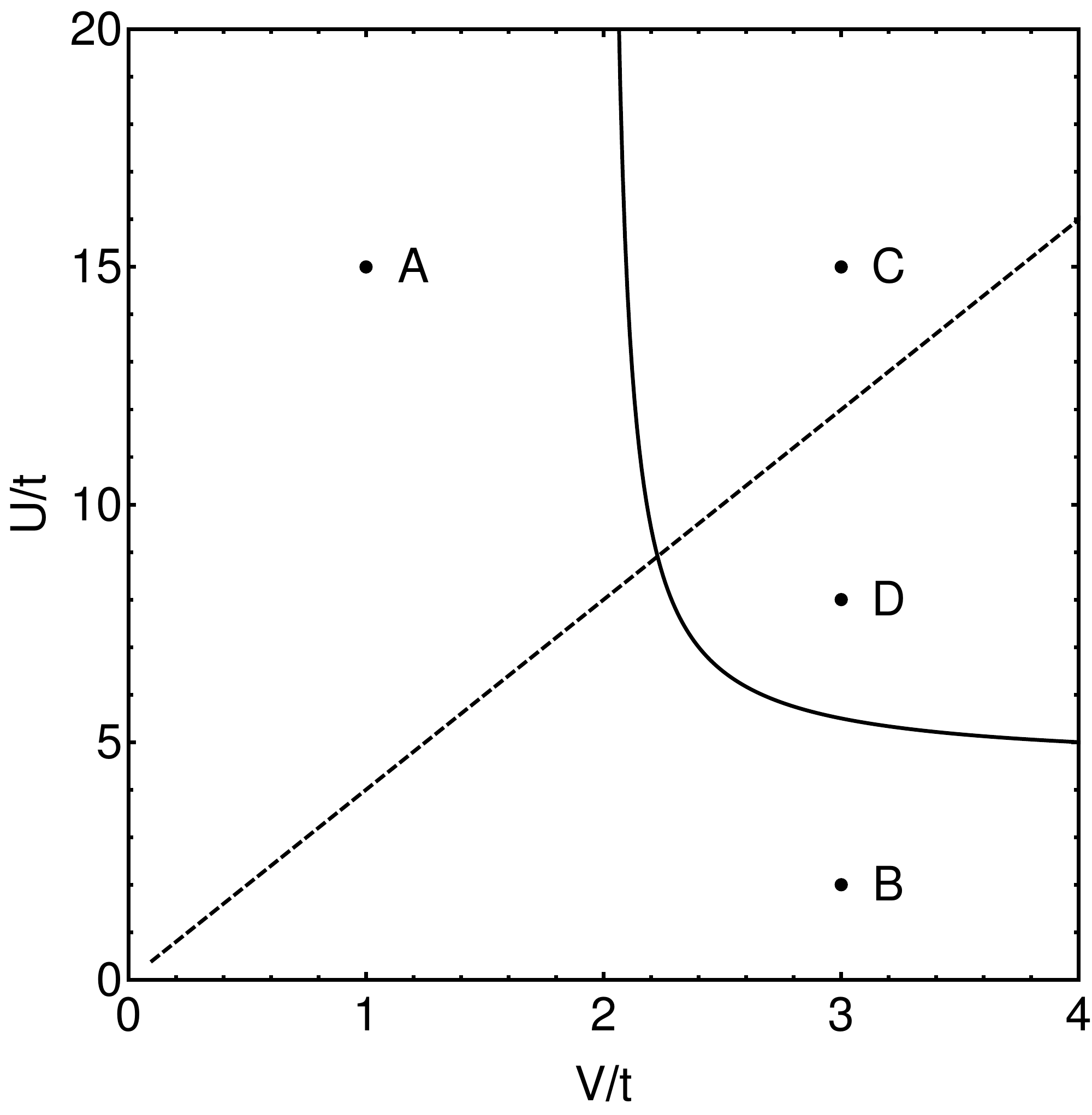}
  \caption{ Phase diagram of the possible structures at the edge for a smooth confining potential ($w = 10$) as a function
  of the interaction strengths $U$ and $V$. A half-filled incompressible plateau exists at all parameters. Additional 
  one and three quarter-filled plateaus (also incompressible) exist roughly in the region above the solid black line. 
  The half-filled plateau features AF (CDW) order in the region above (below) the dashed black line. 
  The points A, B, C, and D mark the parameters $U, V$ used for panels (a), (b), (c), and (d) of Fig.~2 respectively.} 
\end{figure}

\begin{figure*}[t]
  \centering
  \includegraphics[width=0.24\textwidth]{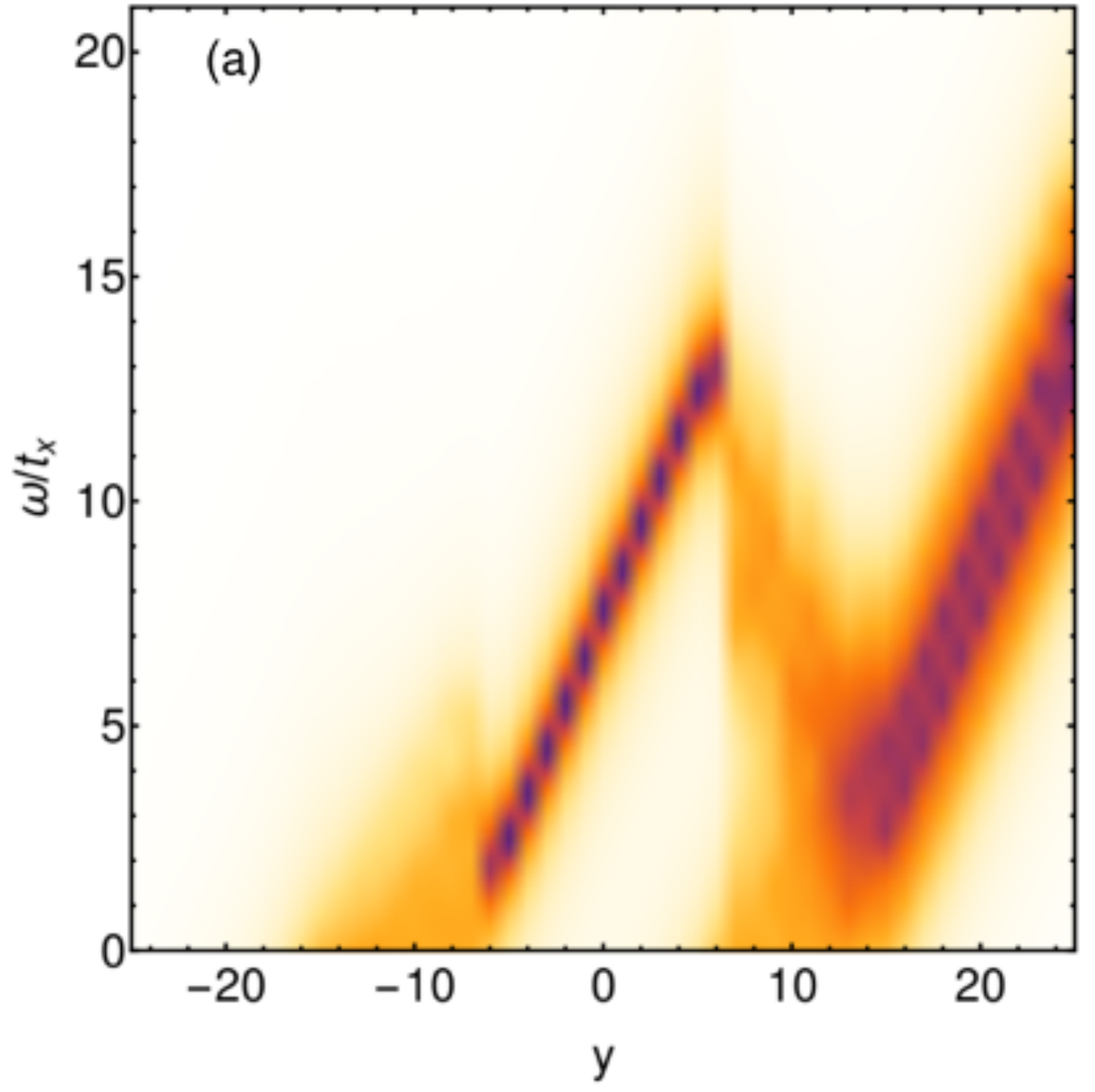}
  \includegraphics[width=0.24\textwidth]{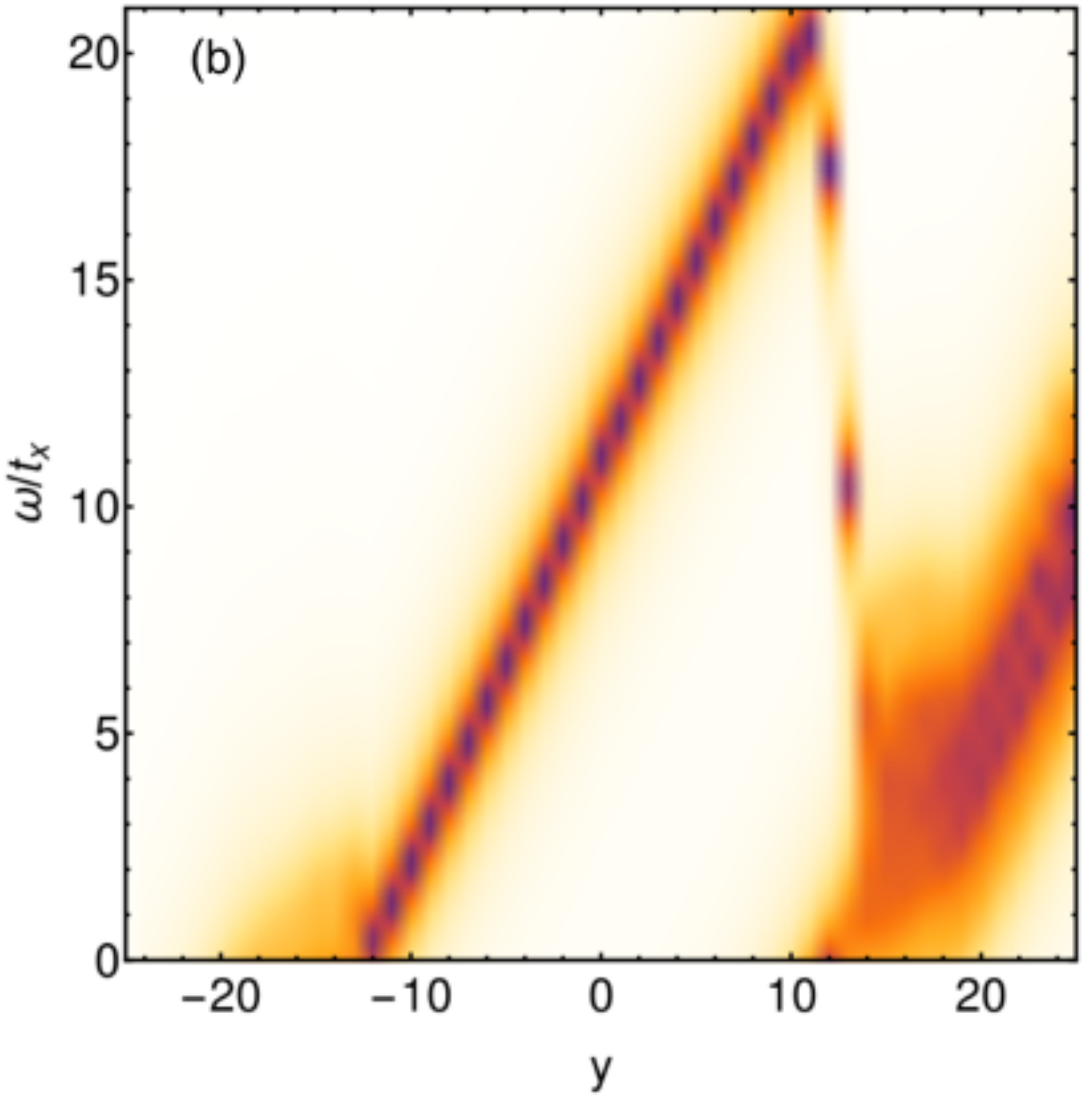}
  \includegraphics[width=0.24\textwidth]{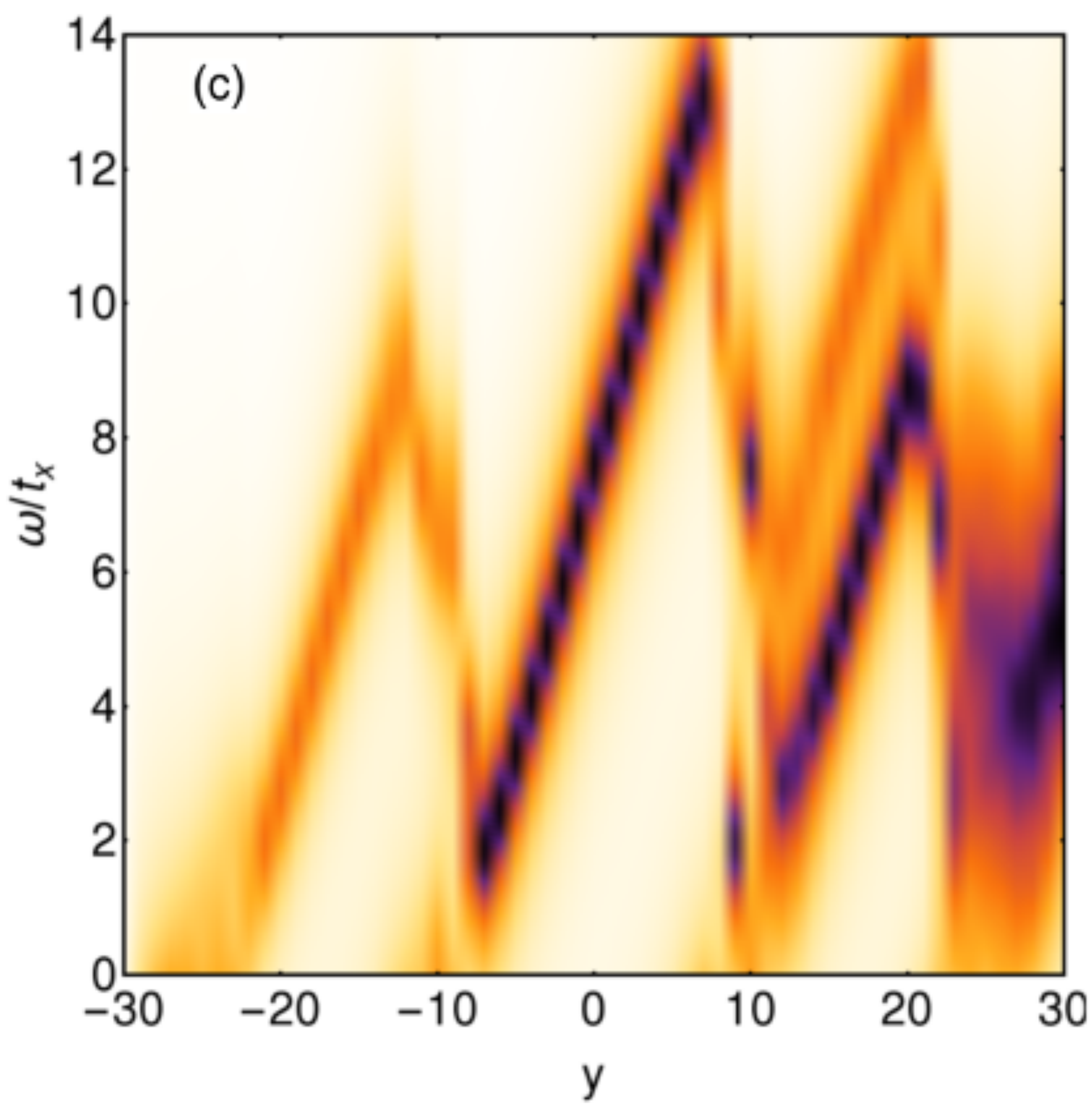}
  \includegraphics[width=0.24\textwidth]{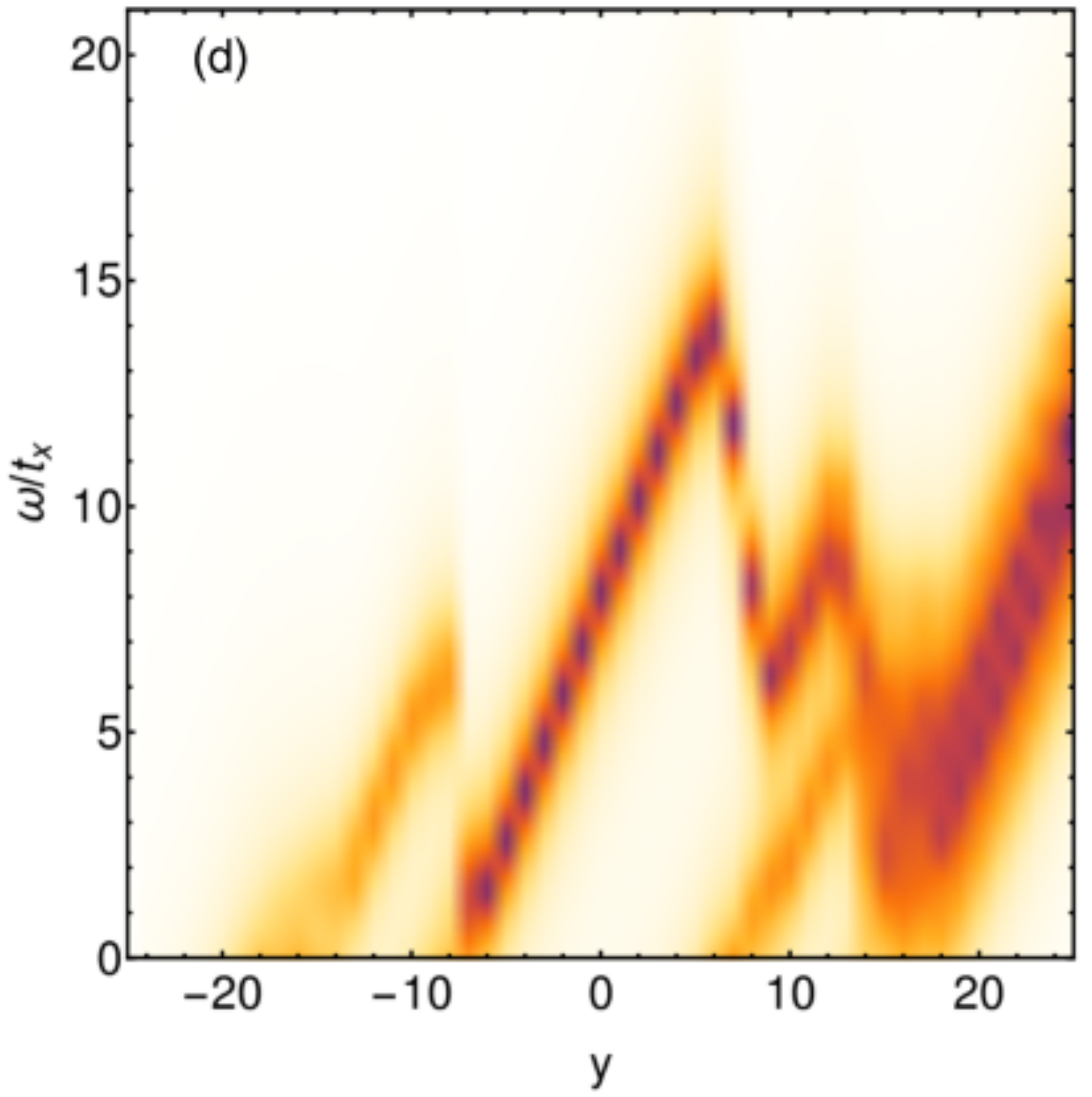}
  \caption{ 
  Tunneling density of states ($\rho_{e}$) in the different structures at the edge as a function of 
  energy $\omega$ and position $y$. The white color denotes zero density. The regions with darker colors 
  have a larger density of states. Panels (a)--(d) correspond to the states shown in Figs.~2(a)--2(d) respectively.
  Comparing with Fig.~2, we see that the incompressible plateaus feature a gap in the tunneling
  density of states, i.e., no states at $\omega \sim 0$. By contrast, the compressible regions feature a 
  finite density of states even at $\omega \approx 0$. Therefore the incompressible (compressible) regions are
  expected to be insulating (metallic) in a clean system. }
\end{figure*}

States which maintain translation invariance along $x$ can only have a nonzero $\Delta$ for $q = 0$. 
By allowing averages with $q \neq 0$ we include states that break this symmetry, through periodic 
(charge and spin) density waves, within the family of allowed variational states. Furthermore, treating $\Delta(q)$ at each $y$ as an
independent parameter allows for states with different kinds of density waves in different regions (along $y$). 
In this Letter, we restrict $q$ to $0, \pi$, and $\pi/2$, which allow for density waves with 
periodicity 1, 2, and 4 along the $x$ direction. 
We have checked that including other values of $q$ (such as $\pi/4$) does not change our results. However, it is possible that larger systems
may exhibit density waves with longer periods. 

We characterize the ground state through the (quantum) expectation value of the charge ($\langle n_{i,y} \rangle $) and spin 
($\langle \vec{s}_{i,y} \rangle$) at each site in the $x$-$y$ plane
(the $x$ coordinate of $\vsr$ is set to an integer $i$).
We define the $y$-dependent average density [$\rho(y)$], as well as the F and
AF order parameters: $\vec{s}_F (y) = \sum_{i} \langle \vec{s}_{i, y} \rangle / L_{x} $ 
and $\vec{s}_{AF} (y) = \sum_{i} (-1)^{i} \langle \vec{s}_{i, y} \rangle / L_x$
in terms of the expectation values. The ground state can be classified into different phases depending on the 
behavior of these order parameters as a function of $y$ (also discretized to the index of the chains). 
To further probe the nature of the ground state we compute the local compressibility  
$\kappa (y) = \partial \rho(y) /\partial \mu$. Finally, we compute the electron 
tunneling density of states $\rho_{e} (\omega, y) =  \text{ Im} G(\omega, y) / \pi$ where $G(\omega, y) = \sum_{n} \sum_{i}
\frac{\psi_{n}^{*}(i, y) \psi_{n} (i, y)}{\omega - \epsilon_{n} - i 0^{+}}$ is the local single-particle Green's function,
and $\epsilon_{n}$ ($\psi_{n}$) denote the energy (wave function) of the $n{\text{th}}$ self-consistent single-particle
state.

{\it Results and discussion}.--We performed the self-consistent HF analysis described above for system sizes up to $64$ sites along $x$ for each chain 
and $75$ chains along $y$ employing a range of parameters $w$, $U$, and $V$, keeping the tunneling energy 
fixed, $t = 1$~\cite{fN-anisotropy}. For very steep confining potentials, we found (as expected) 
a sharp transition between regions of double and zero occupation [Fig.~1(a)]. On the other hand, for sufficiently smooth 
edge potentials ($w \gtrapprox 5$), we observed at least four different structures (shown in Fig.~2 for $w = 50$) 
depending on the values of the interaction strengths ($U$ and $V$). Figure~3 presents the average occupation and spin 
on each site of the 2D space for the structure shown in Fig.~2(c). Figure~4 depicts the regions in the $U$-$V$ plane 
which show the four structures, at smooth confining potentials ($w = 10$). Note that $w = 12$ ($w = 10$) in Fig.~3 
(Fig.~4), while the parameters $U$, $V$, and $t$ are the same as those employed in Fig.~2. A smaller value of $w$ was employed 
because Figs.~3 and 4 show results for a system with a smaller number of chains (30). 

The black lines in Figs.~2(a)--2(d) show the variation of the average density ($\rho$) at the edge 
in the four phases. The central plateau (at $\rho = 1$) can feature AF [Fig.~2(a)] or CDW [Fig.~2(b)] order. 
AF arises when $U \gg V \gg t$; all the sites on this stripe are singly occupied, and the expectation value of spins
follows AF ordering.  
This can be understood as resulting from the Anderson superexchange interaction~\cite{Anderson50,Anderson61,RMP-2DAFM}.
On the other hand, when $U \ll V$ the central stripe features a period 2 CDW [i.e., average occupation $(2,0,2,0,\dots)$ along $x$] 
with no magnetic structure. The large $V$ prefers empty sites next to doubly occupied sites rather than
singly occupied sites. These two structures of the central stripe appear in all our phases. 
For sharp potentials ($w \sim 0$), the AF to CDW transition occurs at $U = 2V$. The transition point gradually shifts toward 
$U = 4V$ as the confining potential is made smoother and does not vary further for $w \geq 5$. 
As mentioned, when both $U$ and $V$ are large and that potential is sufficiently smooth, 
there appear additional plateaus at $\rho = 1/2, 3/2$ [Figs.~2(c) and 2(d)]. 
These additional plateaus have CDW structures, with occupations $(1,0,1,0,\dots)$, 
so that the density is characterized by a wave vector $(\pi,\pi)$. 
Our self-consistent HF analysis also finds ferromagnetic order in the compressible stripes. 
This may be understood in terms of Stoner ferromagnetism since $U \gg t$~\cite{Stoner}. 
However for strictly 1D systems and screened interactions, the Lieb-Mattis theorem
rules out the possibility of a ferromagnetic ground state~\cite{Lieb-Mattis}. In light of this, we believe that the 
long-range ferromagnetism is an artifact of our approximate analysis, which ignores quantum fluctuations, 
and is unlikely to survive a more thorough treatment.
Previous works on 1D chains and ladders, employing bosonization, report that long-range CDW order may exist 
if the filling is close to commensuration and electronic repulsion
is sufficiently strong~\cite{Giamarchi,CDW-Ladder,CDW-Ladder2}. Therefore we expect that our predictions
on charge density order will not be modified qualitatively even if quantum fluctuations are included.

We find that, for all edge structures, the stripes with plateaus have vanishing compressibility while the regions 
in between feature finite compressibility [Figs.~2(e)--2(h)]. This is consistent with the behavior of the tunneling density of 
states (Fig.~5). 
Specifically, the density of states at low energies vanishes (is finite) in the incompressible (compressible) regions. 
Our results indicate that the compressible regions host gapless single-particle states
which could lead to metallic behavior. By contrast the incompressible regions feature a spectral gap 
and are therefore expected to be insulating. These results are consistent with previous works reporting the concomitant
appearance of metallic behavior and finite compressibility~\cite{CSG_Conductivity,Codben99,Ilani_Compressibility,
Kaan_2003,Armagnat_2020}. Our prediction, the emergence of metallic channels at the edge of a trivial insulator,  
can be verified by dc measurements. The presence of compressible and ordered incompressible stripes 
can also be measured by local STM and AFM experiments. 

The above analysis employed PBC along the edge. 
Using open boundary conditions instead, does not affect the incompressible regions strongly. 
By contrast, within the HF approximation, additional incommensurate density modulations 
($q$ is not a multiple of $2\pi / L_{x}$) 
arise in the compressible regions~\cite{Supplemental}. The amplitude of these modulations decreases as the width of these regions increases. 

{\it Conclusions.}--We have studied edge reconstruction at the boundary of a topologically trivial
insulator at zero magnetic field. Employing the self-consistent Hartree-Fock method we find 
that upon smoothening the confining potential, alternating compressible-incompressible stripes emerge 
at the edge. The incompressible stripes appear in the regions where
the average density is close to commensuration and exhibit charge and/or spin
ordered phases which are qualitatively similar to those found in quasi-1D systems at constant filling. This similarity
of phases suggests that the order found in our approximate analysis may survive quantum fluctuations.
Our findings of compressible-incompressible stripes represent a novel metal-insulator transition 
at the edge, which is driven by the interplay of the confining potential, the kinetic energy, and the 
electronic interactions. Our results set the stage for a future detailed investigation of the effects of
quantum fluctuations and disorder on these striped phases.

\begin{acknowledgments}
We acknowledge illuminating discussions with Ganpathy Murthy, Bernd Rosenow, and Niels John. 
Y.G. was supported by DFG RO 2247/11-1, MI 658/10-1, and CRC 183 (Project No.~C01), the Minerva Foundation, 
the German Israeli Foundation (Grant No. I-118-303.1-2018), the Helmholtz International Fellow Award, and by the 
Italia-Israel QUANTRA grant. O.E.W. and A.A. acknowledge support by the Pazy Foundation.
\end{acknowledgments}

%% file: Supplemental.tex
\setcounter{affil}{0}
\setcounter{page}{1}
\renewcommand{\thefigure}{S\arabic{figure}}
\setcounter{figure}{0}
\renewcommand{\theequation}{S\arabic{equation}}
\setcounter{equation}{0}
\renewcommand\thesection{S\arabic{section}}
\setcounter{section}{0}

\title{Supplemental Material for ``Edge Reconstruction of a Time-Reversal Invariant Insulator: Compressible-Incompressible Stripes''}

\author{Udit Khanna}
\affiliation{Department of Condensed Matter Physics, Weizmann Institute of Science, Rehovot 76100, Israel}
\affiliation{Present Affiliation: Department of Physics, Bar-Ilan University, Ramat Gan 52900, Israel}

\author{Yuval Gefen}
\affiliation{Department of Condensed Matter Physics, Weizmann Institute of Science, Rehovot 76100, Israel}

\author{Ora Entin-Wohlman}
\affiliation{School of Physics and Astronomy, Tel Aviv University, Tel Aviv 6997801, Israel}

\author{Amnon Aharony}
\affiliation{School of Physics and Astronomy, Tel Aviv University, Tel Aviv 6997801, Israel}

\begin{abstract}

This supplemental material provides details regarding our Hartree-Fock analysis.

\end{abstract}

\maketitle

\begin{centering}
\subsection*{1. Basic Setup}
\end{centering}

\noindent
To describe the edge of a 2D insulator, we consider the single band extended Hubbard model 
on a square lattice (with lattice spacing 1) with a confining potential perpendicular to it 
(along $y$). The Hamiltonian of the system is $H = H_{o} + H_{ee}$ where the one body term is,
\begin{align}
  H_{o} = &\sum_{\vsr, \sigma} [-\mu + V_{c}(y)] c^{\dagger}_{\vsr, \sigma} c^{}_{\vsr, \sigma} \nonumber \\
  &-t \sum_{\langle \vsr, \vR \rangle, \sigma} \big[ c^{\dagger}_{\vsr, \sigma} c^{}_{\vR, \sigma} + \, \text{h.c.} \big]. 
\end{align}
Here $\vR$ represents only nearest neighbors of $\vsr$, $\sigma$ are the spin indices (quantized along an arbitrary direction), 
$t$ is the hopping integral, and $\mu$ is the chemical potential. We model the confining potential as,
\begin{align}
  V_{c} (y)= \Bigg\{\begin{array}{cc}
    0 & y < -w   \\
    V_{0} + \frac{V_{0}}{w} y & y \geq -w \end{array}
\end{align}
where $V_{0}$ is chosen such that the average occupation of the chain at $y = 0$ is equal to $1$. 
The parameter $w$ is the inverse slope of the edge potential. 

The two-body term ($H_{ee}$) is,
\begin{align}
  H_{ee} = \sum_{\vsr} U n_{\vsr, \ua} n_{\vsr, \da} + \sum_{\langle \vsr, \vR \rangle} V n_{\vsr} n_{\vR}
\end{align}
where $n^{}_{\vsr, \sigma} = c^{\dagger}_{\vsr, \sigma} c^{}_{\vsr, \sigma}$ is the (spin-resolved) number operator, and
$n^{}_{\vsr} = \sum_{\sigma} n^{}_{\vsr, \sigma}$ is the total number at a site. The onsite
interaction strength is denoted $U$ while the nearest neighbor repulsion is denoted as $V$.
We assume periodic boundary conditions along the edge (along $x$). 
Using $c_{\vsr \sigma} = \sum_{k} c_{k y \sigma} e^{i k x} / \sqrt{L_x}$, the single-particle Hamiltonian 
changes to,
\begin{align}
  H_{o} = &\sum_{k, y, \sigma} [-\mu + V_{c}(y) - 2t \cos (k)] c^{\dagger}_{k, y, \sigma} c^{}_{k, y, \sigma} \nonumber \\
  &-t \sum_{\langle y, y^{\prime} \rangle, k, \sigma} \big[ c^{\dagger}_{k, y, \sigma} c^{}_{k, y^{\prime}, \sigma} + \, \text{h.c.} \big]
\end{align}
and the two-body term becomes,
\begin{align}
  H_{ee} &= \sum_{y k_1 k_2 q} \big[ U + 2V\cos(q) \big] 
  c^{\dagger}_{k_1, y, \ua} c^{\dagger}_{k_2, y, \da} c^{}_{k_2 - q, y, \da} c^{}_{k_1 + q, y, \ua} \nonumber \\
  &+ \sum_{y k_1 k_2 q \sigma} V \cos(q) c^{\dagger}_{k_1, y, \sigma} c^{\dagger}_{k_2, y, \sigma} 
  c^{}_{k_2-q, y, \sigma} c^{}_{k_1+q, y, \sigma} \nonumber \\ 
  &+ \sum_{y k_1 k_2 q \sigma \sigma^{\prime}} V c^{\dagger}_{k_1, y, \sigma} c^{\dagger}_{k_2, y+1, \sigma^{\prime}} 
  c^{}_{k_2 - q, y+1, \sigma^{\prime}} c^{}_{k_1+q, y, \sigma}
\end{align} 

For the results presented in the main text, we used $\mu = 2U + 12V$, which is sufficiently large for each site to 
be doubly occupied in the absence of the confining potential $V_c$. $V_{0}$ was set to $3U/2 + 8V$. \\

\begin{centering}
\subsection*{2. Hartree-Fock Analysis}
\end{centering}

\begin{figure*}[t]
  \centering
  \includegraphics[width=0.34\textwidth]{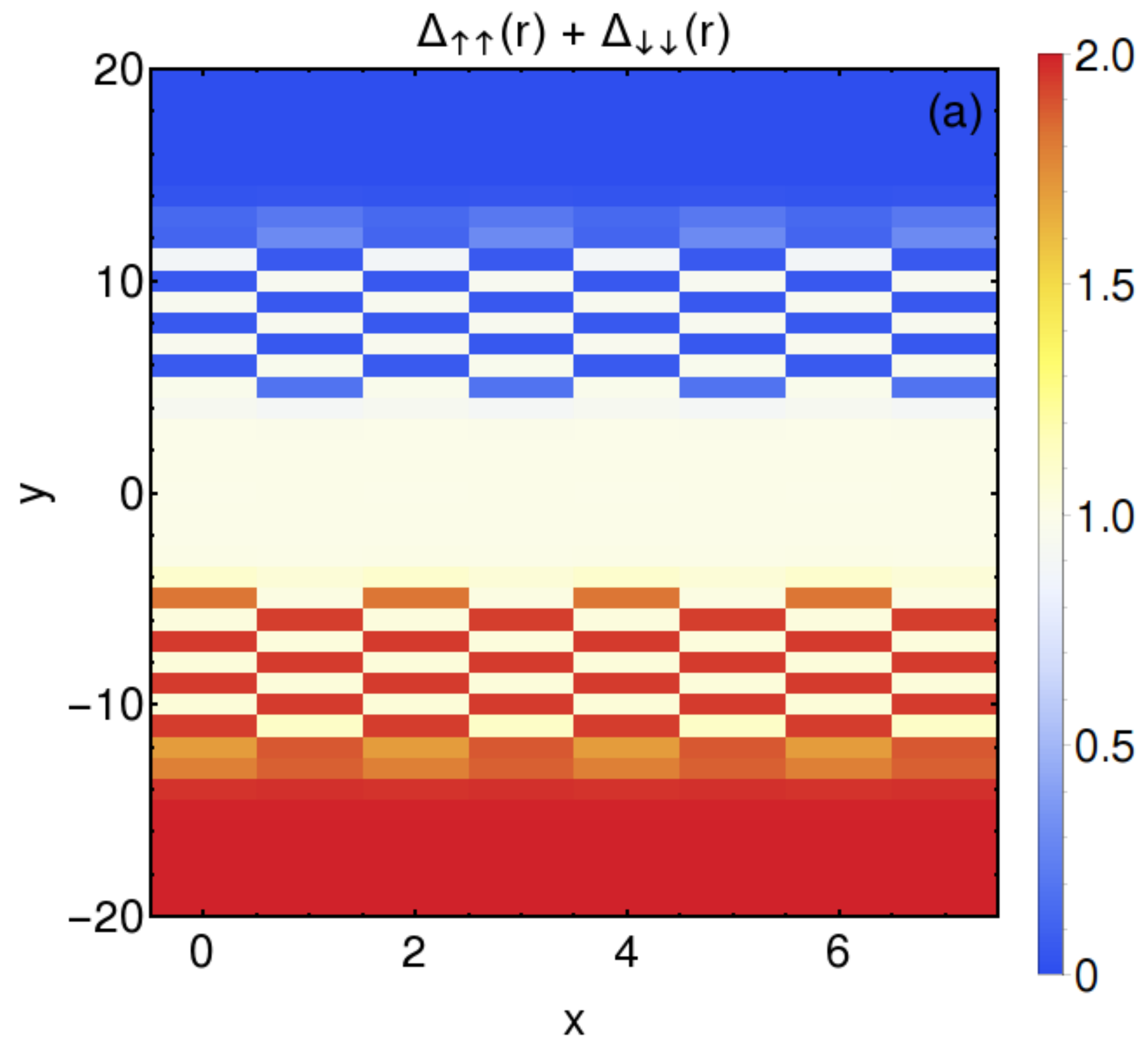}
  \includegraphics[width=0.34\textwidth]{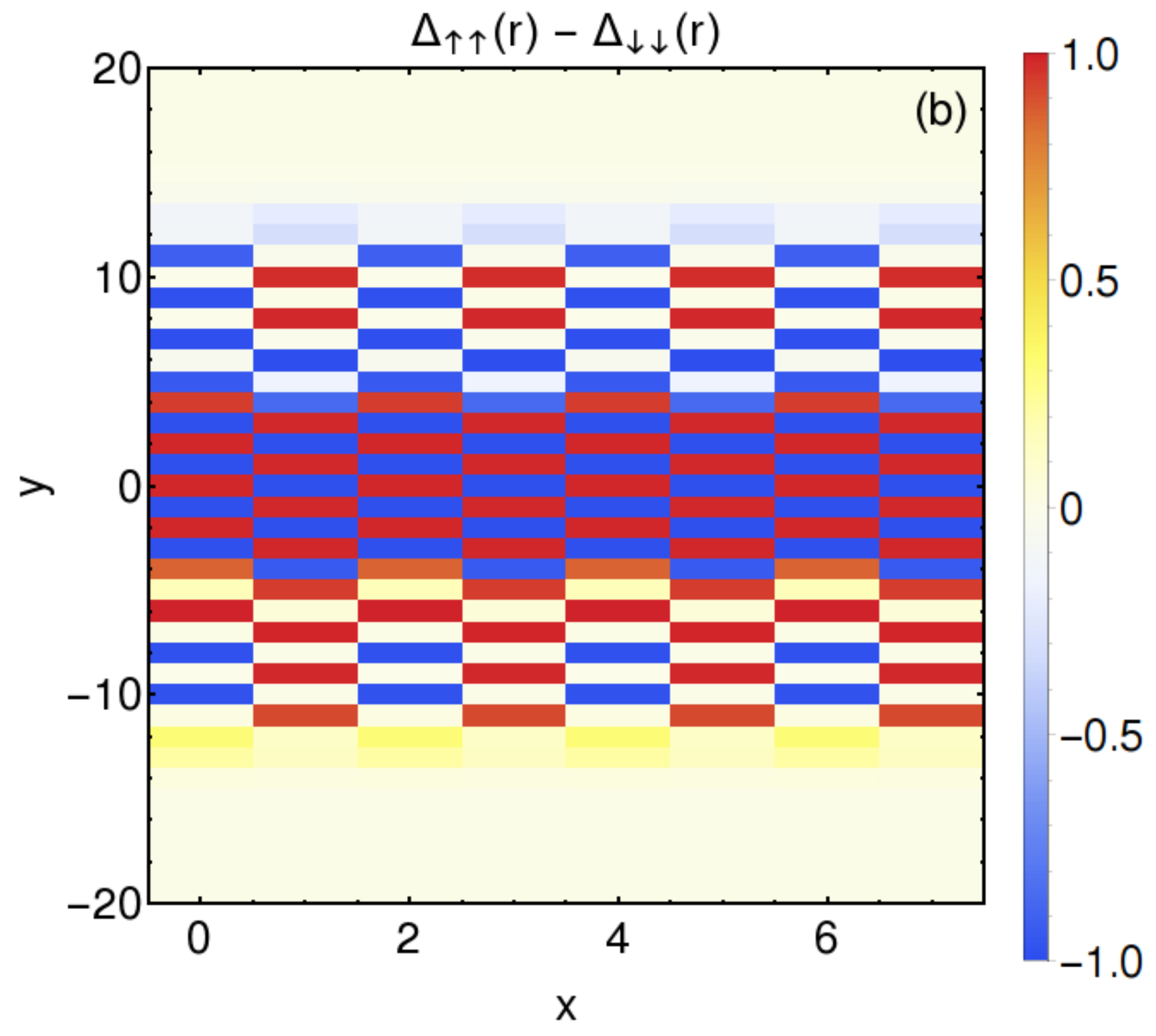}
  \includegraphics[width=0.34\textwidth]{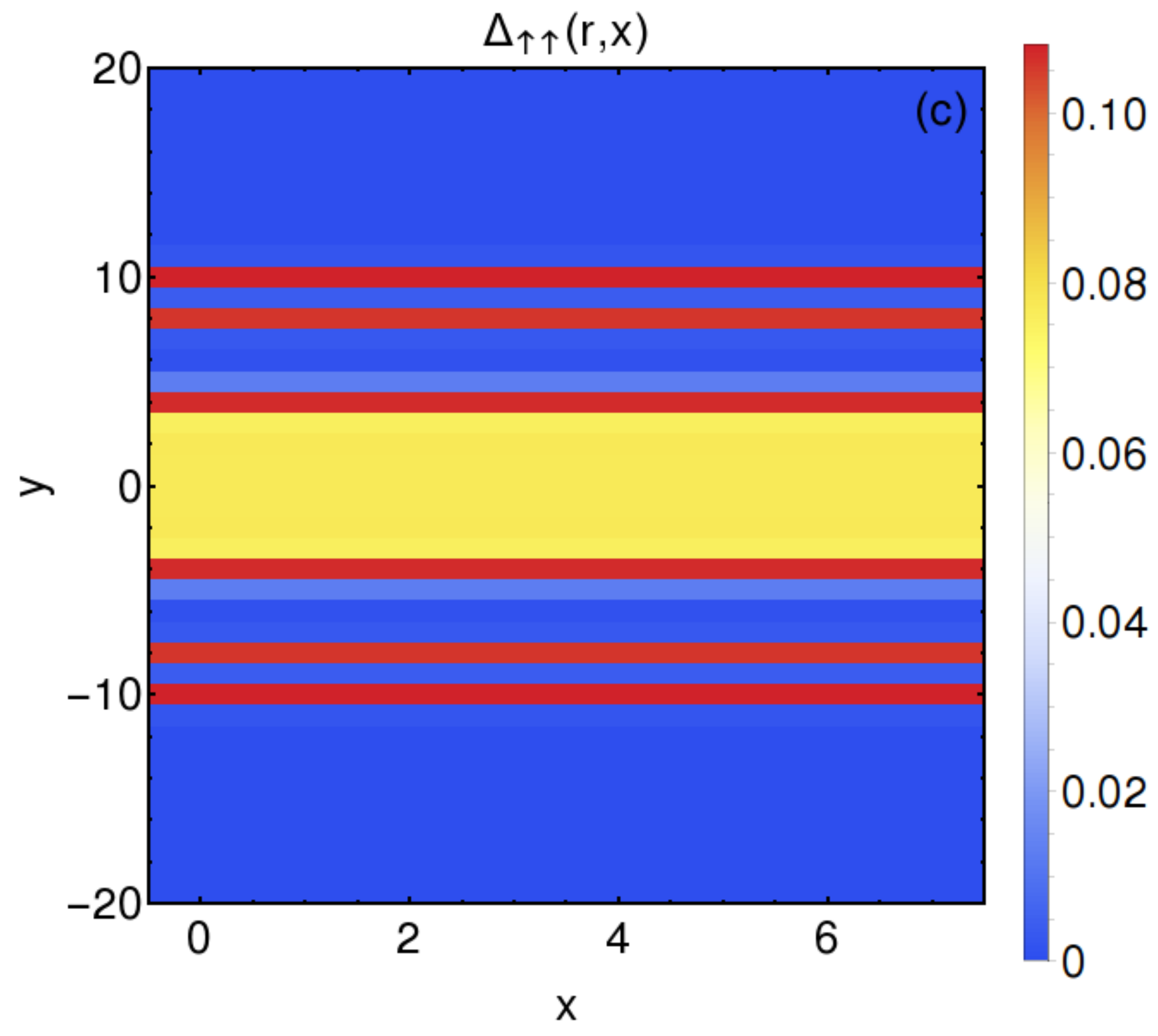}
  \includegraphics[width=0.34\textwidth]{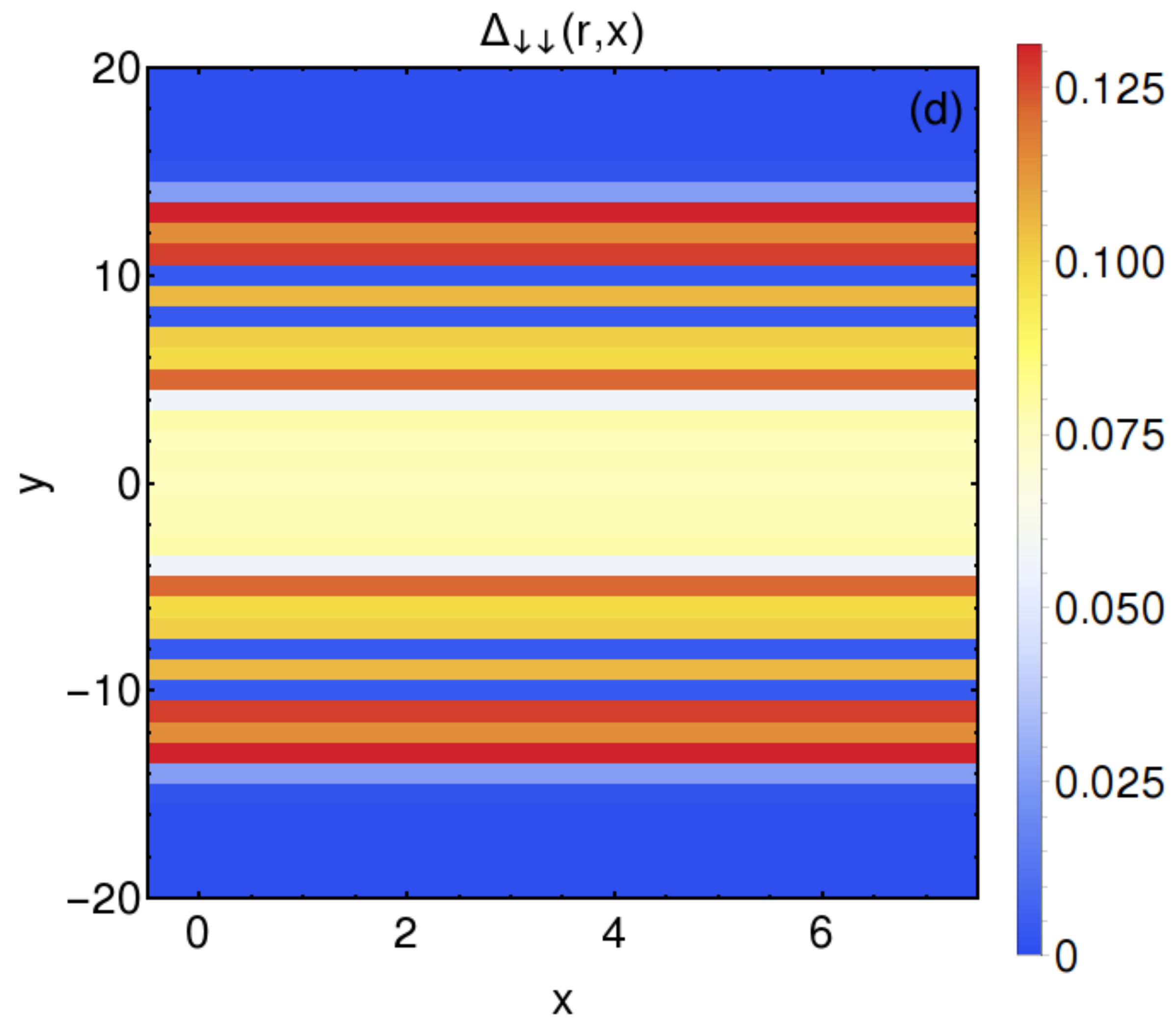}
  \includegraphics[width=0.34\textwidth]{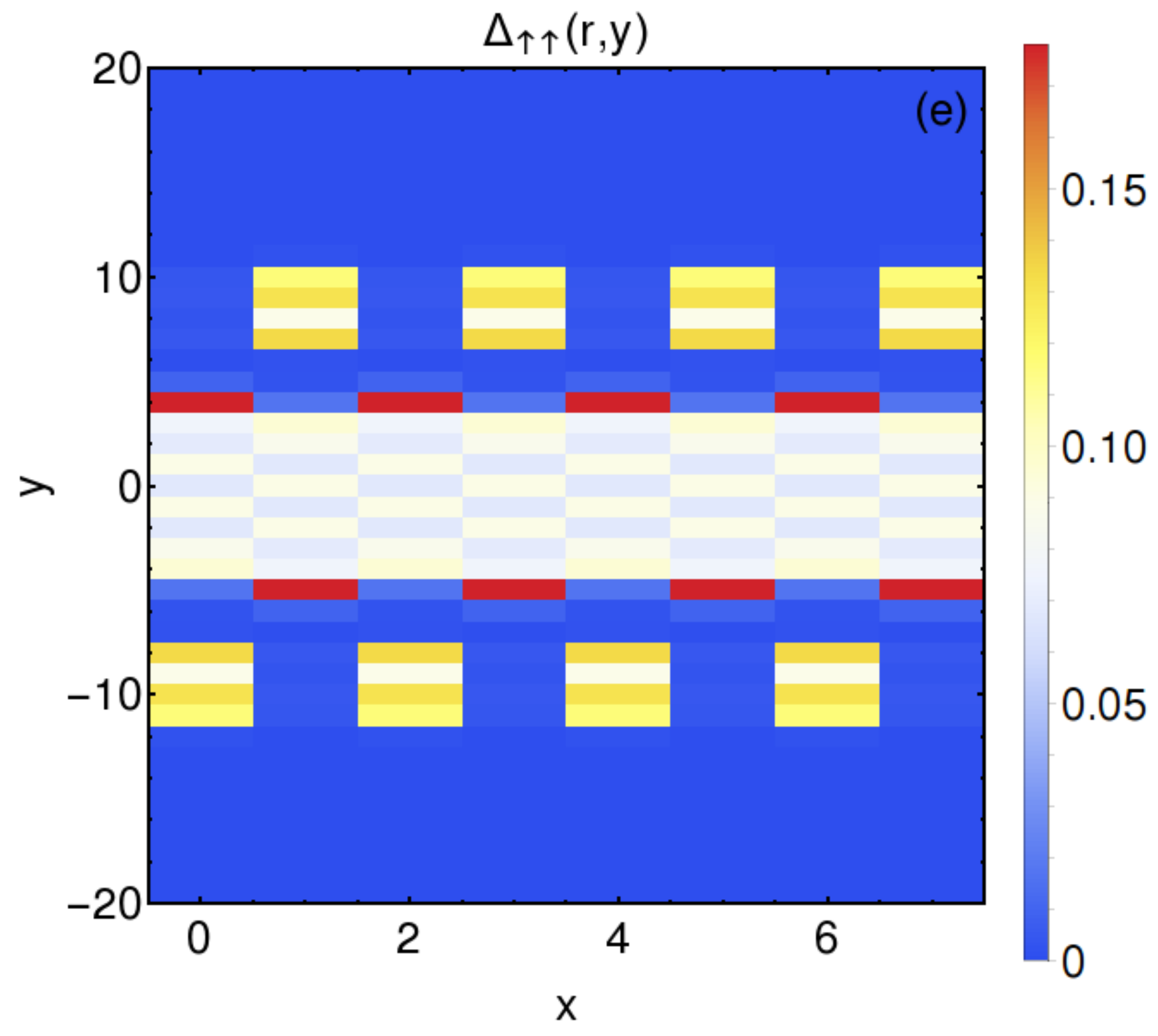}
  \includegraphics[width=0.34\textwidth]{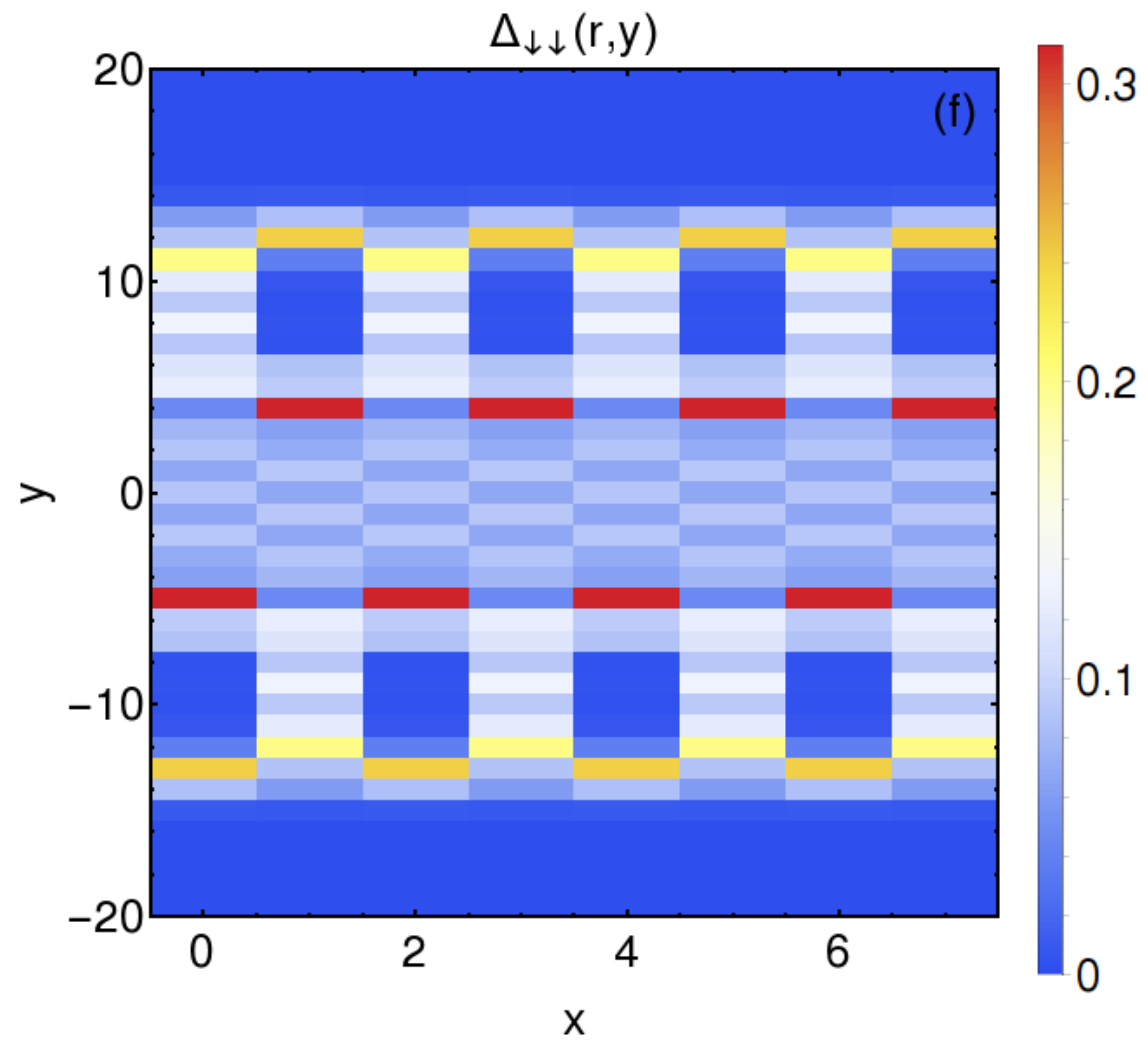}
  \caption{ 
  Self-consistent single-particle averages ($\Delta$) for the edge structure shown in Fig.~2(c) of the main text. Panels (a,b) depict 
  combinations of the onsite averages. Panels (c,d) [(e,f)] show the absolute value of the averages between nearest neighbors 
  along $x$ [$y$] direction. Only the first 8 sites along the edge ($x$) are shown here. 
  The parameters $U, V, t$ used here are the same as those used for Fig.~2(c). 
  However the value of $w$ was reduced to $30$ because only 40 chains were included in this calculation. }
\end{figure*}

\begin{figure*}[t]
  \centering
  \includegraphics[width=0.34\textwidth]{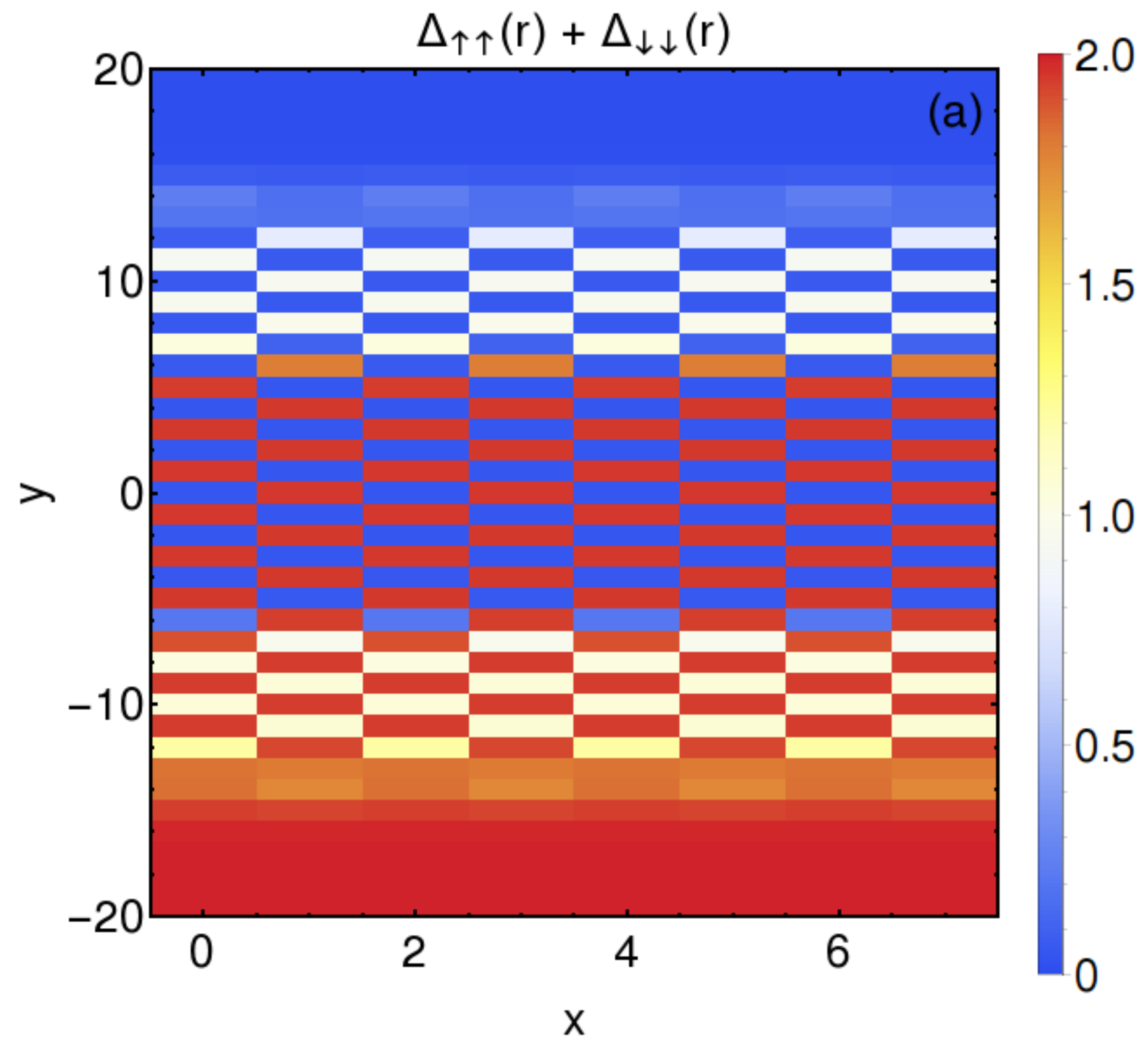}
  \includegraphics[width=0.34\textwidth]{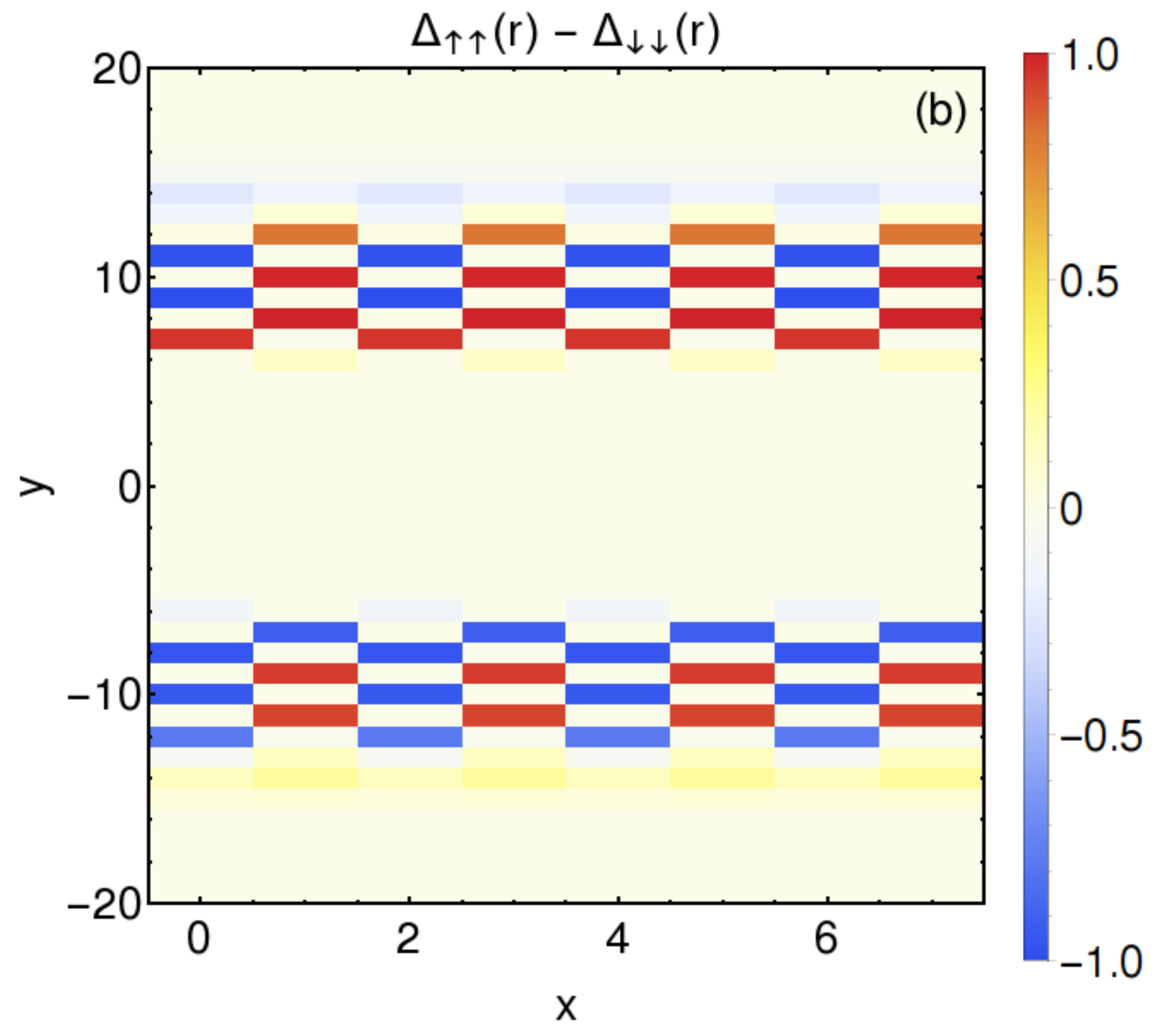}
  \includegraphics[width=0.34\textwidth]{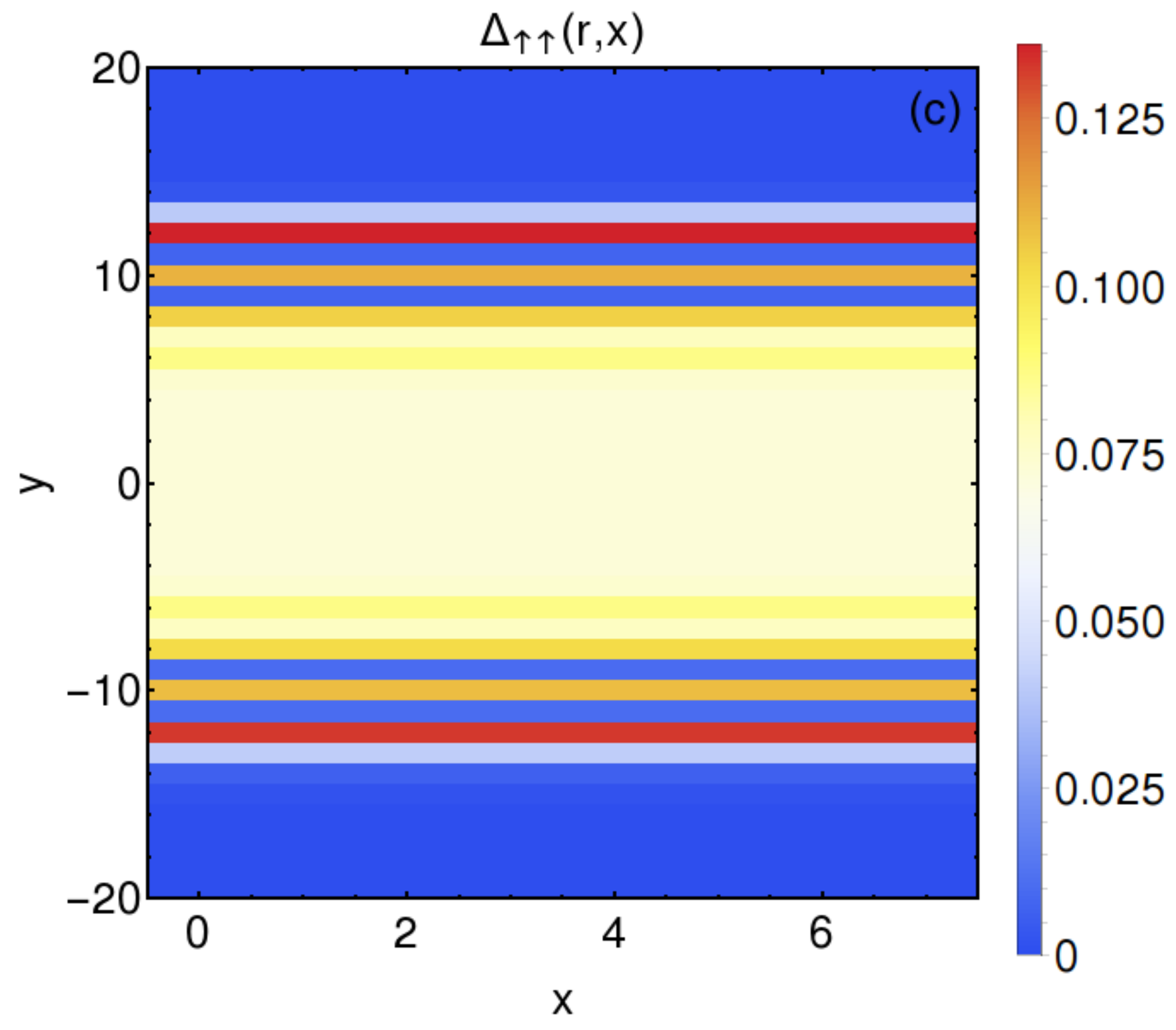}
  \includegraphics[width=0.34\textwidth]{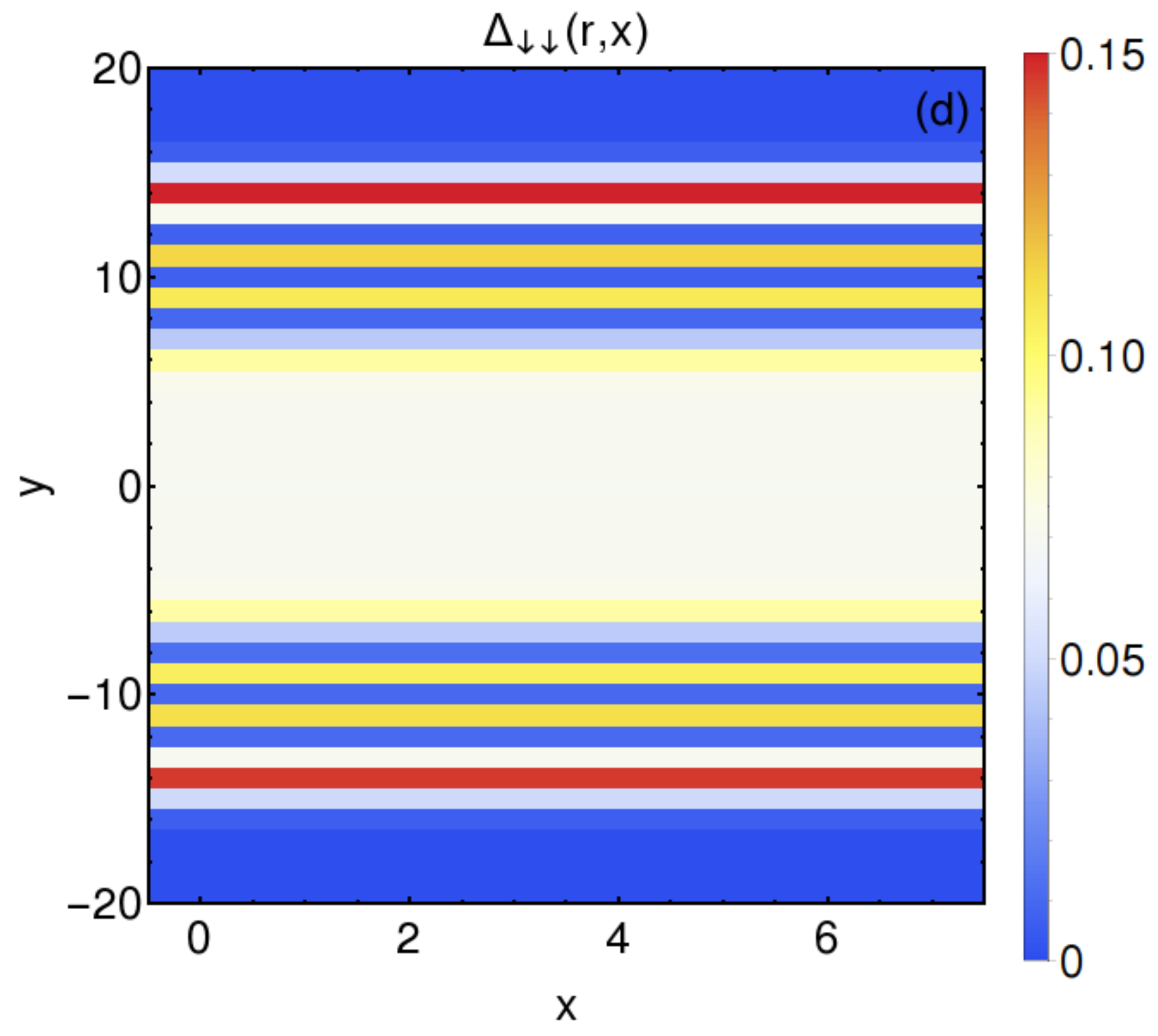}
  \includegraphics[width=0.34\textwidth]{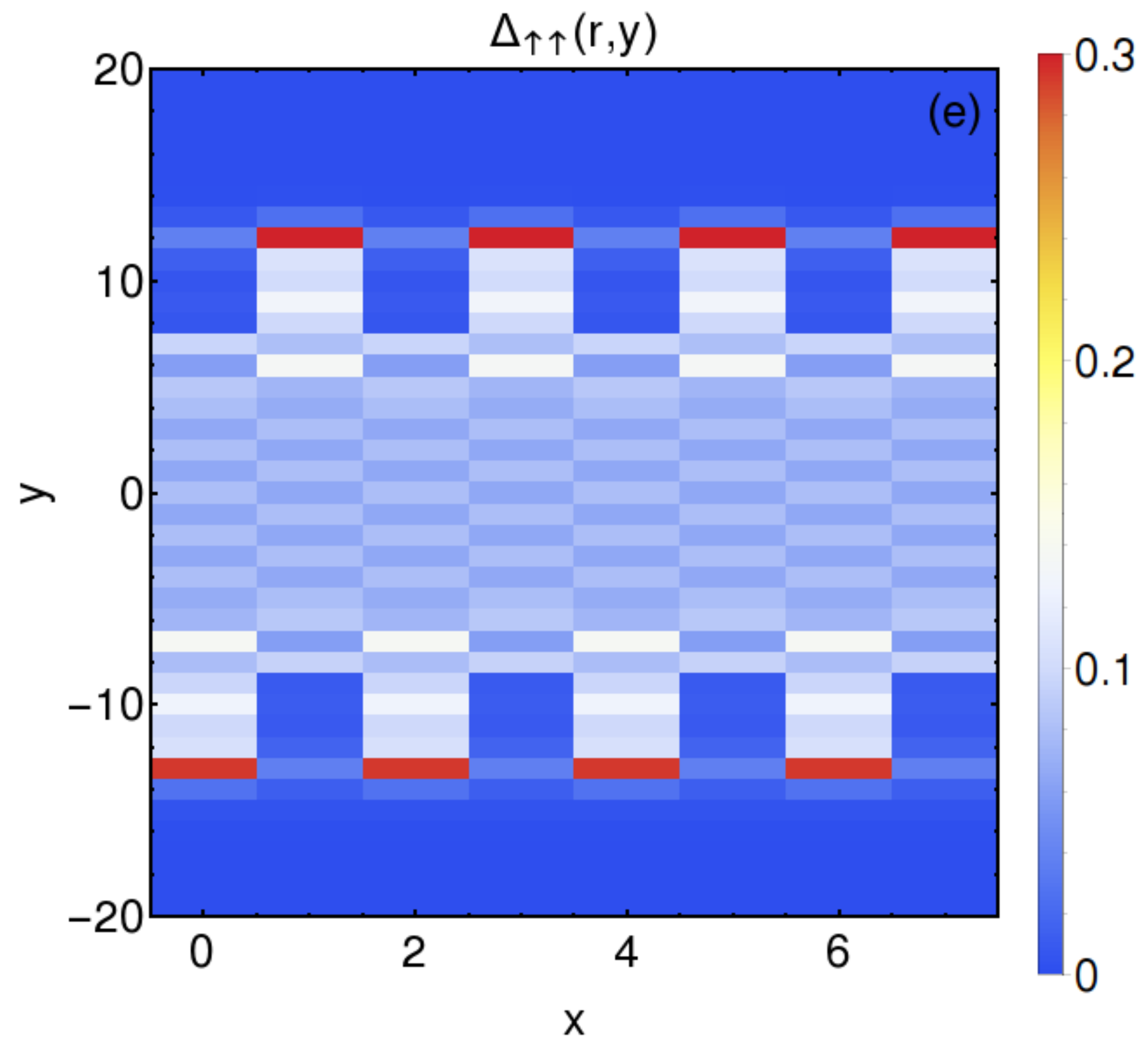}
  \includegraphics[width=0.34\textwidth]{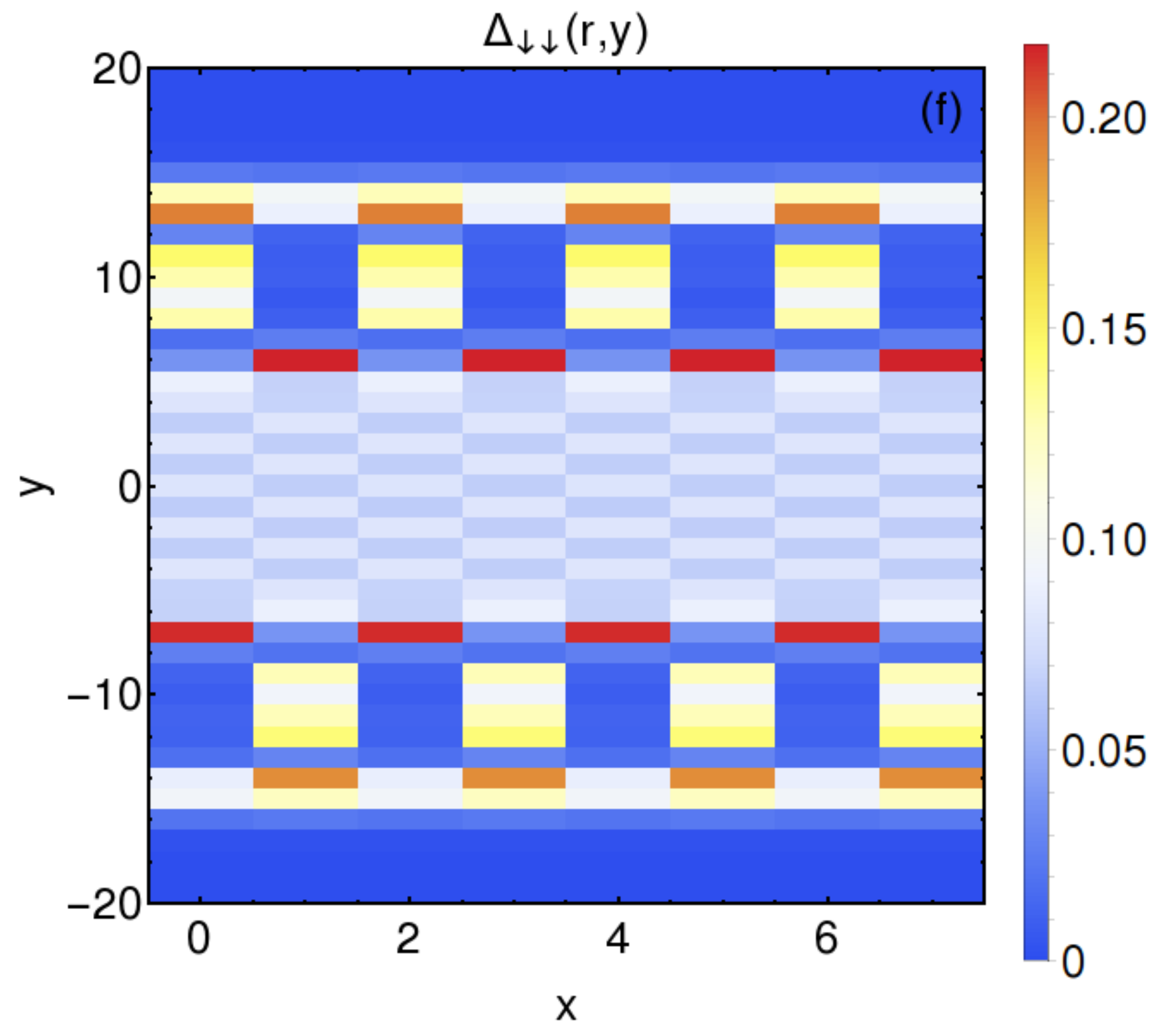}
  \caption{ 
  Self-consistent single-particle averages ($\Delta$) for the edge structure shown in Fig.~2(d) of the main text. Panels (a,b) depict 
  combinations of the onsite averages. Panels (c,d) [(e,f)] show the absolute value of the averages between nearest neighbors 
  along $x$ [$y$] direction. Only the first 8 sites along the edge ($x$) are shown here. 
  The parameters $U, V, t$ used here are the same as those used for Fig.~2(d). 
  However the value of $w$ was reduced to $30$ because only 40 chains were included in this calculation. }
\end{figure*}

The two-body term ($H_{ee}$) is treated within the Hartree-Fock (HF) approximation. Defining the single particle 
averages through,
\begin{align}
  \Delta_{\sigma \sigma^{\prime}} (k, q, y, \delta) = \langle c^{\dagger}_{k, y, \sigma} c^{}_{k+q, y+\delta, \sigma^{\prime}} \rangle
\end{align}
the HF approximation of $H_{ee}$ is,
\begin{align}
  \nonumber
   H_{ee} &\approx  H_{\text{HF}} = \sum_{k q y \sigma} \mathcal{J}_{c} (k, q, y, \sigma) 
   \cdo{k, y, \sigma} \co{k+q, y, \sigma} \nonumber \\
   &+ \mathcal{J}_{f} (k, q, y, \sigma) \cdo{k, y, \sigma} \co{k+q, y, \bar{\sigma}} \nonumber \\
   &+ \Big[ \mathcal{K} (k, q, y, \sigma, \sigma^{\prime}) \cdo{k, y, \sigma} \co{k+q, y+1, \sigma^{\prime}} 
   + \text{h.c.} \Big] \label{S7}
\end{align}
where, $\mathcal{J}_{c/f}$ and $\mathcal{K}$ are the self-consistent one-body (spin-dependent) potentials and tunneling amplitudes 
generated in the mean-field approximation. In terms of the single-particle averages these are,
\begin{align}
  \mathcal{J}_{c} &= \frac{1}{L_x} \sum_{p} \bigg\{
    \big[ U + 2 V \cos(q)\big] \Delta_{\bar{\sigma} \bar{\sigma}} (p, -q, y, 0) \nonumber \\ 
     &+ 2 V \cos(q) \Delta_{\sigma \sigma} (p, -q, y, 0) \nonumber \\ 
     &- 2 V \cos(p - k) \Delta_{\sigma \sigma} (p + q, -q, y, 0) \nonumber \\ 
     &+ V \sum_{\sigma^{\prime}, \alpha = \pm 1} \Delta_{\sigma^{\prime} \sigma^{\prime}} (p, -q, y+\alpha, 0)
     \bigg\} 
   \label{S8}
\end{align}
\begin{align}
  \mathcal{J}_{f} &= -\frac{1}{L_x} \sum_{p}  
    \big[ U + 2 V \cos(p - k)\big] \Delta_{\bar{\sigma} \sigma} (p + q, -q, y, 0) ,
    \label{S9}
\end{align}
and
\begin{align}
  \mathcal{K} &= - \frac{1}{L_x} \sum_{p} V \Delta_{\sigma^{\prime} \sigma} (p, -q, y + 1, -1) .
  \label{S10}
\end{align}
Note that the dependencies of $\mathcal{J}_{c/f}$ and $\mathcal{K}$ [specified in Eq.~(\ref{S7})] were suppressed 
in Eqs.~(\ref{S8}-\ref{S10}) above for brevity.

Since $H_{o} + H_{HF}$ is a single-particle Hamiltonian, it can be easily diagonalized numerically
given the values of $\mathcal{J}$ and $\mathcal{K}$. The many-body ground state of $H$ is approximated 
by the self-consistent Slater determinant ground state of $H_{o} + H_{HF}$. Here self-consistency implies that 
the single-particle averages ($\Delta$) used to define $H_{HF}$, through Eqs.~(\ref{S8}-\ref{S10}), are equal to those 
evaluated in the lowest energy Slater determinant state of $H_{o} + H_{HF}$. 

In order to find the self-consistent state, we implement an iterative procedure. Each iteration starts from 
a diagonalization of the approximate Hamiltonian for a given set of single-particle averages ($\Delta$). 
The many-body ground state is defined as the Slater determinant state for which 
all single-particle levels with energies less (greater) than the chemical potential are 
occupied (empty). This ground state is then used to compute new single-particle averages, which are then
used in the next iteration. 
The process is repeated until the largest (absolute) change in the averages at the end of each step becomes smaller than a 
given cutoff (chosen to be $10^{-10}$ here). The iteration is initialized using random values for the single-particle
averages, and the entire iteration is repeated multiple times employing different initial conditions. 
We have confirmed that all initial conditions end up with the same solution for the self-consistent 
state (upto global spin-rotations and time-reversal operations). 

The self-consistent $k$-space single-particle averages [$\Delta_{\sigma \sigma^{\prime}} (k, q, y, \delta)$] evaluated 
at the end of the iteration are then used to compute single-particle averages in real space as follows,
\begin{align}
  \langle c^{\dagger}_{x_{i} y \sigma} &c^{}_{x_{j} y^{\prime} \sigma^{\prime}} \rangle 
  = 1/L_{x} \sum_{k, k^{\prime}} \langle c^{\dagger}_{k y \sigma} 
  c^{}_{k^{\prime} y^{\prime} \sigma^{\prime}} \rangle e^{i (k^{\prime} x_{j} - k x_{i})} \nonumber \\
  &= 1/L_{x} \sum_{k, q} \langle c^{\dagger}_{k y \sigma} 
  c^{}_{k+q y^{\prime} \sigma^{\prime}} \rangle e^{i q x_{j}} e^{i k (x_{j} - x_{i})} \nonumber \\
  &= 1/L_{x} \sum_{k, q} \Delta_{} (k, q, y, \delta) e^{i q x_{j}} e^{i k (x_{j} - x_{i})} 
\end{align}
In the last equality above, we used $y^{\prime} = y + \delta$. The onsite averages ($x_{i} = x_{j}$, 
$\delta = 0$) may then be used to evaluate the expectation values of density ($\rho$) and spin ($\vec{s}\,$) as defined
in the main text. Similarly we may find the nearest neighbor averages along $x$ ($x_{j} = x_{i} + 1$, $\delta = 0$) and
along $y$ ($x_{j} = x_{i}$, $\delta = 1$). 
Figures~S1, S2 depict the real space onsite [in panels (a,b)] and nearest neighbor [along $x$ in (c,d) 
and along $y$ in (e,f)] averages for the structures shown in Fig.~2(c,d) of the main text. 

Note that we restrict ourselves to $q = 0, \pi$ and $\pi/2$, which allow for density waves with
periodicity 1, 2 and 4 along the $x$ direction. This is because we had first performed real space HF 
analysis (with smaller system sizes) and only found solutions with these periodicities. However we have 
checked that our results, based on the $k$-space HF analysis, do not change even if the system size 
is increased and other values of $q$ (such as $\pi/4$) are included.  \\

\begin{figure}[t]
  \centering
  \includegraphics[width=0.34\textwidth]{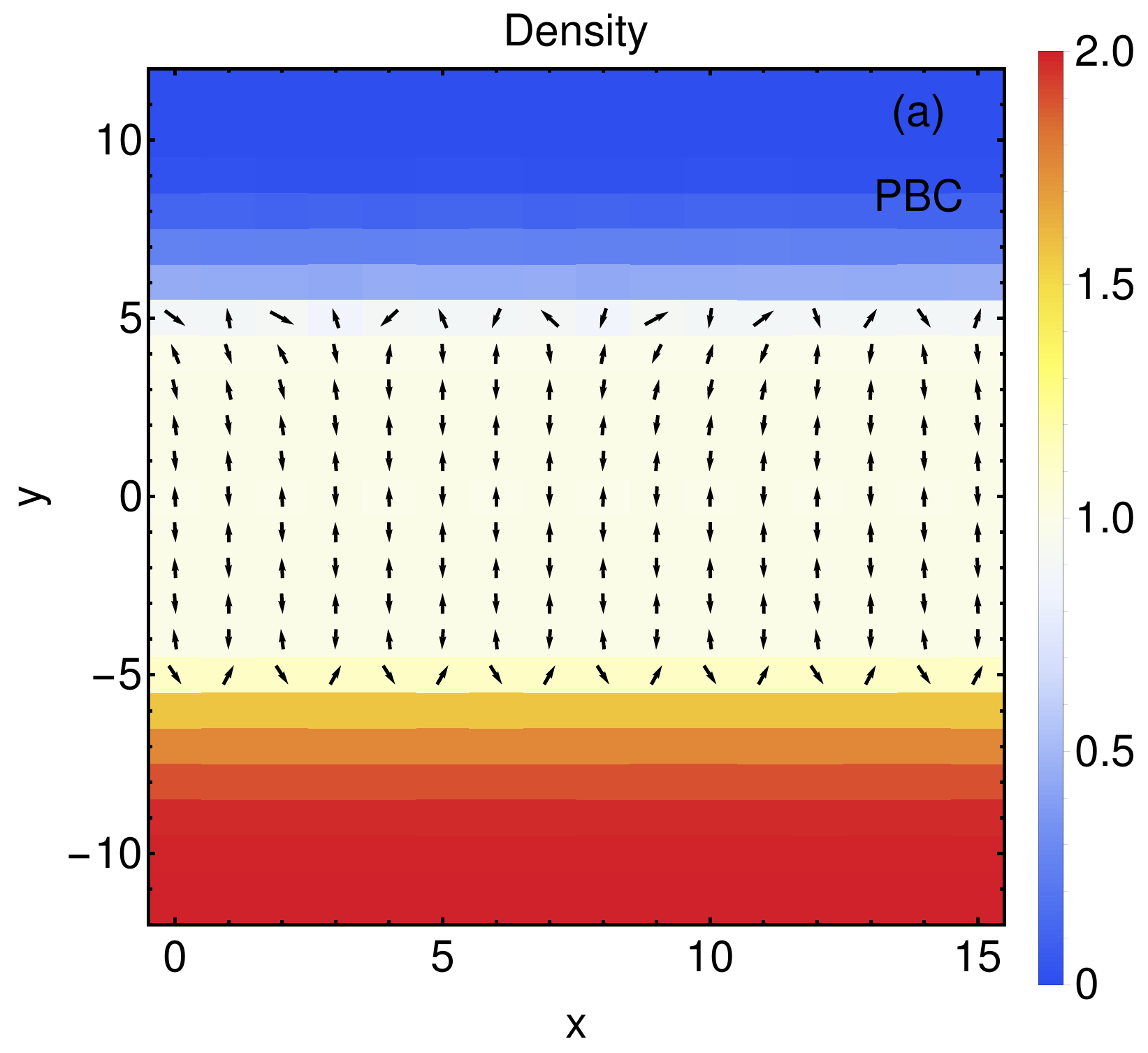}
  \includegraphics[width=0.34\textwidth]{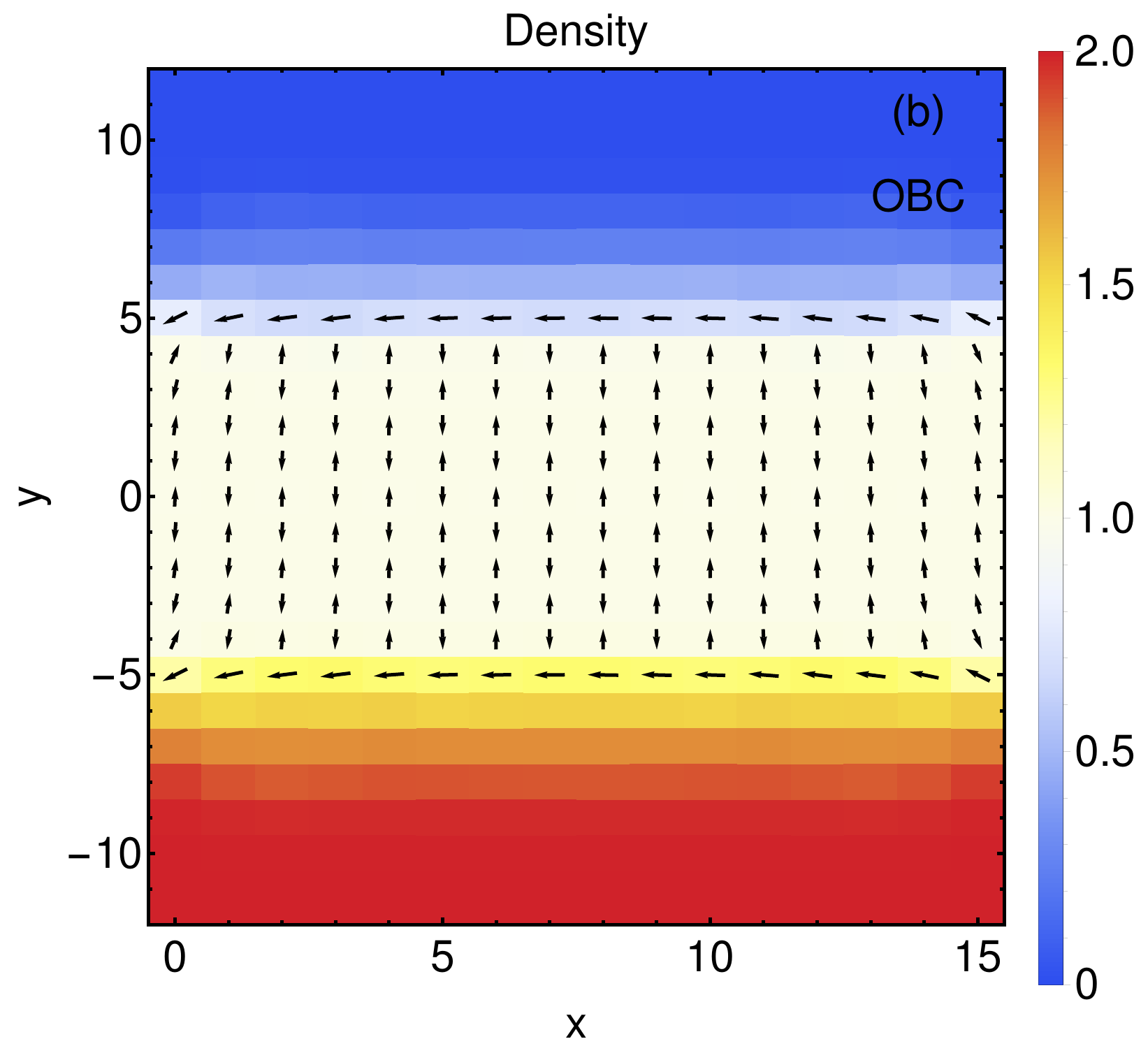}
  \caption{ 
  Comparison of HF ground states with (a) periodic and (b) open boundary conditions along the edge ($x$) for small systems 
  ($L_{x} = 16$, $L_{y} = 25$). 
  The color map shows the average occupation at each site. The black arrows on the sites represent the average spin 
  (along $s_z$ direction) at each site.  
  The parameters $U, V, t$ used here are the same as those used for Fig.~2(a).  
  However the value of $w$ was reduced to $15$ because only 25 chains were included in this calculation. }
\end{figure}

\begin{centering}
  \subsection*{3. Effect of Boundary Conditions Along the Edge}
\end{centering}

  We used periodic boundary conditions (PBC) along the edge (along the $x$-direction) for the results presented in the main text. 
  On a finite lattice these boundary conditions only allow density waves that are commensurate with the system size 
  ($q$ must be a multiple of $2\pi/L_{x}$). Here, we describe the effect of using open boundary conditions (OBC), which also allow 
  states with incommensurate density waves. In general, we do not expect boundary conditions to strongly influence the 
  incompressible regions due to the gap in local tunnelling density of states. However, this may not be obvious for the gapless 
  compressible stripes (particularly in the quasi-1D limit). 

\begin{figure}[t]
  \centering
  \includegraphics[width=0.34\textwidth]{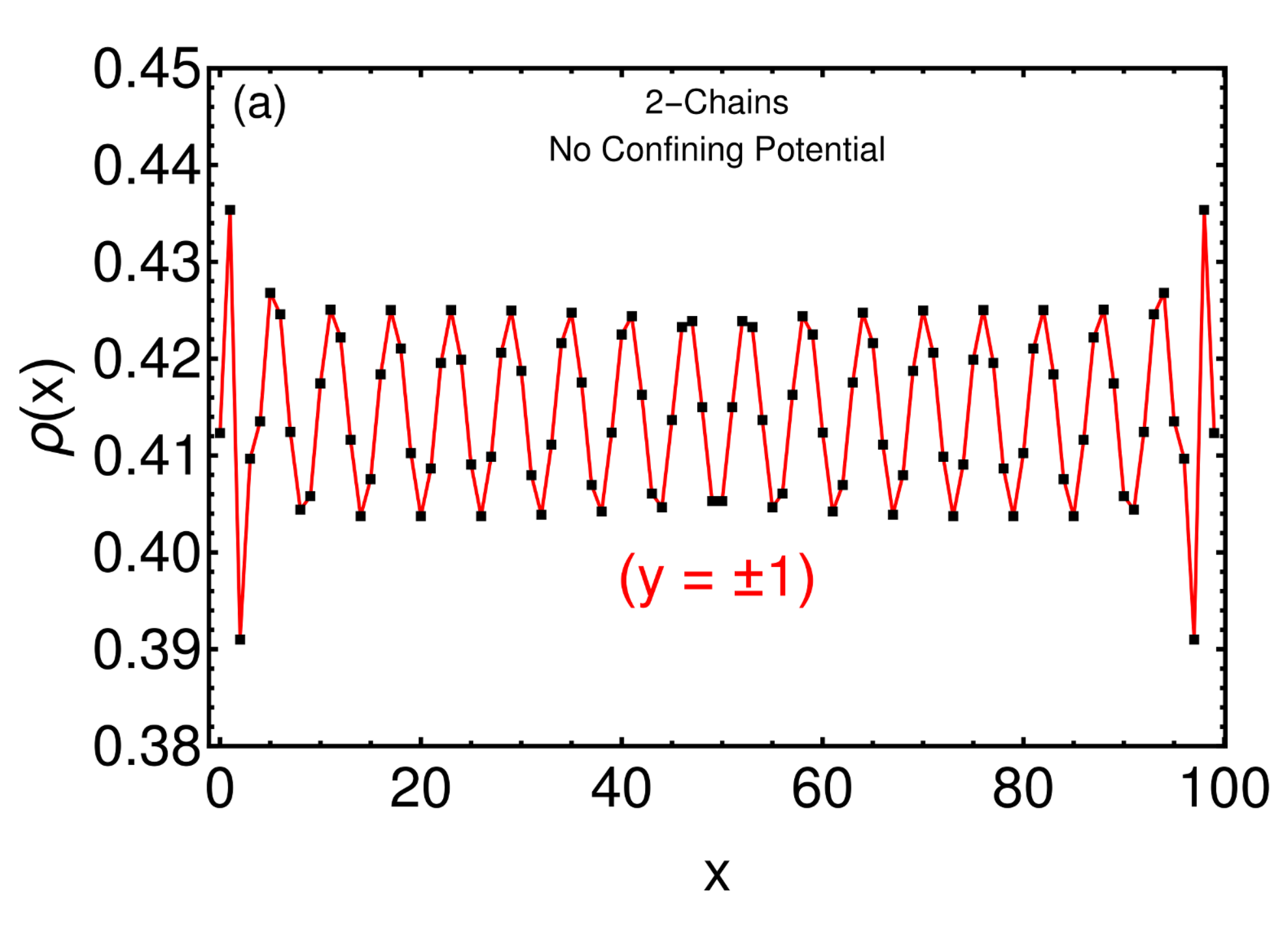}
  \includegraphics[width=0.34\textwidth]{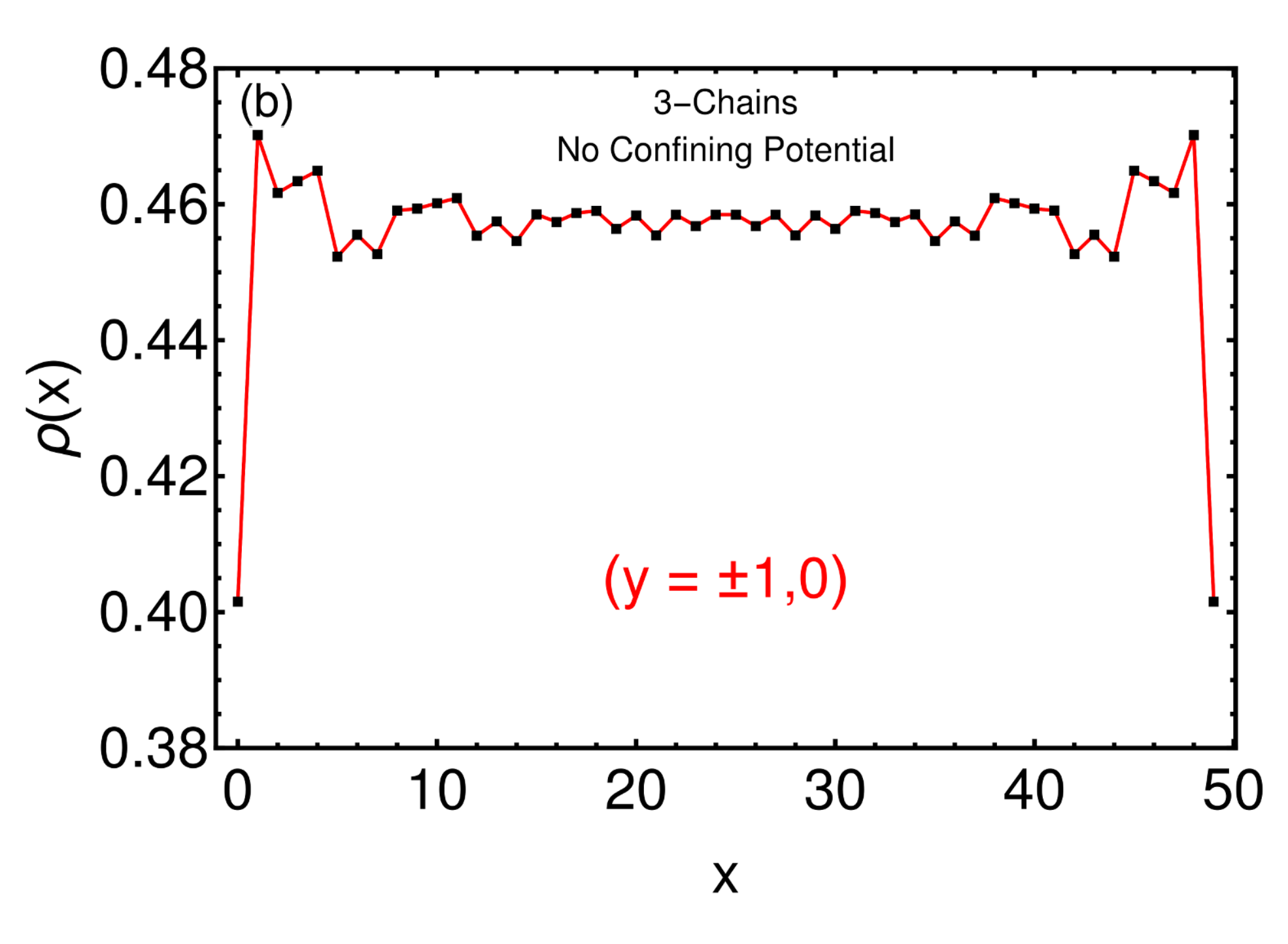}
  \caption{
  Variation of the average occupation along $x$ (along the edge) in systems with (a) 2 and (b) 3 chains. Open boundary condition was 
  assumed along $x$ and the chemical potential was chosen such that the average charge density is away from commensurate 
  values ($\rho = 1$ or $1/2$). Away from the boundaries, an incommensurate charge density wave is evidently present in the case of 2 chains, 
  but is suppressed in the case of 3 chains. These results are in the absence of a confining potential, but do not change 
  qualitatively when a small potential is included. }
\end{figure}

  OBC can be used only in a real-space HF analysis, which is not practical for large systems 
  comprising several chains. Figure~S3 presents a comparison of the HF ground states with different boundary conditions for small systems 
  ($L_{x} = 16$, $L_{y} = 25$). 
  Clearly, the results are almost indentical expect for minor variations of the average spin 
  at boundaries of the incompressible region centered about $y = 0$. This is consistent with our general expectation regarding the
  incompressible regions. However, such an analysis may not be satisfying regarding the fate of the compressible stripes due to their 
  small width (4 or 5 chains) and finite length (16 sites). In order to overcome these size restrictions, 
  we also studied longer (along $x$) systems with a very small number of coupled chains. The global chemical potential is then chosen such 
  that the average charge density of chains is away from commensurate values (1 or 1/2). Such a small system may be expected to mimic a part 
  of the emergent compressible stripes (surrounded by insulating incompressible regions) at the edge of a band insulator. 

  Figure~S4 depicts the variation of the average occupation along $x$ in such small systems. The density varies along $x$ only due to 
  the use of OBC (using PBC instead would result in a uniform density along $x$). 
  Clearly, an incommensurate charge density wave forms away from the boundaries in the case of 2 chains. Our analysis shows that the
  periodicity of this density wave varies continuously with the average density of the chains (tuned through the global chemical potential). 
  In this sense, the system remains compressible irrespective of the boundary conditions used. However, the tunnelling density of states
  reveals that (contrary to the case of PBC) there is a {\it finite gap} in this case. Therefore, (within our HF analysis) this system 
  is compressible but not metallic. On the other hand, both the amplitude of the density modulation and the spectral gap are suppressed in
  the case of 3 chains. We have found that this `even-odd' effect persists for systems with larger number of chains. 
  Furthermore, the presence of a small confining potential along $y$ does not affect these results qualitatively.

  However, we believe that some of the effects of using OBC is an artifact of the HF approximation. The Mermin-Wagner theorem ensures 
  that the long-range incommensurate density wave order may not exist in quasi one-dimensional systems. Therefore, after quantum 
  fluctuations are included, the compressible regions would not support any long-range order 
  irrespective of the number of chains. Our HF analysis using PBC is thus more reliable since it ignores the possibility of such 
  spurious long-range order from the beginning.
  
  The metallicity of compressible regions is more subtle. HF analysis using PBC (OBC) predicts that such regions are gapless (gapped). 
  Additionally, our analysis of systems with a small number of chains (with OBC) shows that the 
  gap in the tunelling density of states generally reduces as the number of chains is increased from 2 to 4. Extrapolating to a 
  large number of chains, it is possible that this trend continues and eventually the gap becomes unobservably small. 
  This supports our {\it conjecture} that the compressible regions found in our analysis are also metallic. 
  The subtlety arises from the presence of quantum fluctuations and the confining potential. 
  The melting of the density-wave order (due to quantum fluctuations) may turn the small energy gap into a {\it soft} gap. However, 
  we emphasise that this expectation is based on studies of systems without any confining potential. The situation is obviously
  different at the edge of a 2D insulator. In addition to a spatially varying potential (along the width of the stripe), the 
  compressible regions are enclosed by two inequivalent incompressible regions. The combined effect of this modification of boundary conditions 
  (along $y$) and quantum fluctuations on the conductance of the compressible regions is an interesting direction for a future study. \\

\begin{centering}
  \subsection*{4. Effective $r_{S}$ of the stripes}
\end{centering}

  The HF ground state comprises of both compressible and incompressible regions. Intuitively, we expect that the kinetic energy 
  dominates of the interaction energy in regions of finite compressibility, and vice versa for the incompressible stripes.
  However, this is not clear from the bare values of the Hamitonian parameters since we used $U, V \gg t$ in our analysis. 
  Therefore we computed an effective $r_{S}$ at each $y$, and compared the values in the compressible and incompressible regions. 

  We defined $r_{S}$ as the ratio of the interaction energy (defined as the expectation value of the two body-terms acting on a given $y$, 
  or $\langle H_{ee} (y) \rangle$) to the absolute value of the single-particle energy (defined as $\langle H_{o} (y) \rangle$). 
  To treat the non-local terms along $y$, we performed an average over the two nearest neighbors along $y$. For instance, the single-particle 
  hopping term along $y$ was treated as $\sim \langle (c^{\dagger}_{x,y,\sigma} c^{ }_{x,y+1,\sigma} + 
  c^{\dagger}_{x,y,\sigma} c^{ }_{x,y-1,\sigma})\rangle/2 + \text{c.c}$, and similarly for $H_{ee}$. 

  We find that for a fixed $U = 15 t$ and varying $V$ (while remaining in the region of phase A), 
  the largest value of $r_{S}$ (as defined here) varies from $\sim 4$ to $6$ for the incompressible region at half-filling ($\rho = 1$),
  and from $\sim 0.2$ to $0.4$ for the compressible regions. We have checked that these values do not depend sensitively on the 
  boundary conditions employed along the edge. These results confirm that indeed the ultimate ground state (in HF) is a combination of strongly
  interacting incompressible regions and weakly interacting compressible regions. Additionally, the results also suggest that it would be 
  incorrect to use intuition based on non-interacting one-dimensional systems (in the absence of a confining potential) to draw concrete 
  conclusions about the setup considered here.

%% file: EdgeSpinChains.bbl
\begin{thebibliography}{54}
\bibitem [{fN-()}]{fN-anisotropy}%
  \BibitemOpen
  \href@noop {} {}\bibinfo {note} {Using different tunneling and/or interaction
  strengths in the two direction does not change the results
  qualitatively.}\BibitemShut {Stop}%
\end{thebibliography}
